\documentclass[a4paper,11pt]{article}
\usepackage[T1]{fontenc}      
\usepackage[utf8]{inputenc}   
\usepackage[english]{babel}   

\usepackage{amsmath, amsfonts,amssymb}
\usepackage{bm,bbm}
\usepackage{listings,lstautogobble,longtable}
\usepackage{graphicx}
\usepackage{rotating}
\usepackage{url}
\usepackage{tabularx,tablefootnote,threeparttable}
\usepackage{bigints}
\usepackage{float}
\usepackage[small,bf]{caption2}
\usepackage{etoolbox}
\usepackage{graphicx}
\usepackage{scalerel}
\usepackage{xcolor}
\usepackage{multirow}
\usepackage{centernot}

\newcommand\independent{\protect\mathpalette{\protect\independenT}{\perp}}
\def\independenT#1#2{\mathrel{\rlap{$#1#2$}\mkern2mu{#1#2}}}

\usepackage{geometry}
\definecolor{blucite}{RGB}{12,127,172}
\usepackage[superscript,biblabel]{cite}
\usepackage[colorlinks, citecolor=blucite]{hyperref}

\begin{document}
	
	\title{\huge\textbf{Causal effect of chemotherapy received dose intensity on survival outcome:\\ a retrospective study in osteosarcoma}\\ \vspace{1mm}}
	
	\author{\Large Marta Spreafico$^{1,*}$, \, Francesca Ieva$^{2,3}$, \, Marta Fiocco$^{1,4,5}$\\ \quad\\
		\footnotesize  $^1$Mathematical Institute, Leiden University, Leiden 2333 CA, The Netherlands\\
		\footnotesize $^2$MOX -- Department of Mathematics, Politecnico di Milano, Milan 20133, Italy\\
		\footnotesize  $^3$Health Data Science Center, Human Technopole, Milan 20157, Italy\\
		\footnotesize $^4$Department of Biomedical Data Sciences, Leiden University Medical Center,  Leiden 2333 ZA, The Netherlands\\
		\footnotesize $^5$Trial and Data Center, Princess M\'{a}xima Center for Pediatric Oncology, Utrecht 3584 CS, The Netherlands\\  \quad \\
		*\texttt{m.spreafico@math.leidenuniv.nl} \quad}
	\date{ }
	
	\maketitle

	\begin{abstract}
		\footnotesize
		 \textbf{Background:}   This study aims to analyse the effects of reducing Received Dose Intensity (RDI) in chemotherapy treatment for osteosarcoma patients on their survival by using a novel approach. Previous research has highlighted discrepancies between planned and actual RDI, even among patients randomized to the same treatment regimen. To mitigate toxic side effects, treatment adjustments, such as dose reduction or delayed courses, are necessary. Toxicities are therefore risk factors for mortality and predictors of future exposure levels. Toxicity introduces post-assignment confounding when assessing the causal effect of chemotherapy RDI on survival outcomes, a topic of ongoing debate.
		 
		 \textbf{Methods:} Chemotherapy administration data from BO03 and BO06 Randomized Clinical Trials (RCTs) in ostosarcoma are employed to emulate a target trial with three RDI-based exposure strategies: 1) \textit{standard}, 2) \textit{reduced}, and 3) \textit{highly-reduced} RDI. Investigations are conducted between subgroups of patients characterised by poor or good Histological Responses (HRe), i.e., the strongest known prognostic factor for survival in osteosarcoma. Inverse Probability of Treatment Weighting (IPTW) is first used to transform the original population into a pseudo-population which mimics the target randomized cohort. Then, a Marginal Structural Cox Model with effect modification is employed. Conditional Average Treatment Effects (CATEs) are ultimately measured as the difference between the Restricted Mean Survival Time of \textit{reduced/highly-reduced} RDI strategy and the \textit{standard} one. Confidence Intervals for CATEs are obtained using a novel IPTW-based bootstrap procedure.
		 
		 \textbf{Results:} Significant effect modifications based on HRe were found. Increasing RDI-reductions led to contrasting trends for poor and good responders: the higher the reduction, the better (worsen) was the survival in poor (good) reponders. Due to their intrinsic resistance to chemotherapy, poor reponders could benefit from reduced RDI, with an average gain of 10.2 and 15.4 months at 5-year for \textit{reduced} and \textit{highly-reduced} exposures, respectively.
		 
		 \textbf{Conclusions:} This study introduces a novel approach to (i) comprehensively address the challenges related to the analysis of chemotherapy data, (ii) mitigate the \textit{toxicity-treatment-adjustment} bias, and (iii) repurpose existing RCT data for retrospective analyses extending beyond the original trials' intended scopes.
	\end{abstract}
	
	\vspace{5mm}  \small
	\noindent \textbf{Key-words:} Marginal Structural Cox Models; Inverse Probability of Treatment Weighting; Effect Modification; Target Trial Emulation; Received Dose Intensity; Chemotherapy; Toxicity; Event-Free Survival \normalsize

\section{Background}\label{s:intro}

Osteosarcoma is a rare malignant bone tumor primarily affecting children, adolescents, and young adults, with an annual incidence of 3-4 patients per million \cite{smeland2019}. While multidisciplinary management, including neoadjuvant and adjuvant chemotherapy with aggressive surgical resection \cite{ritter2010}, has improved clinical outcomes, there has been little progress in survival over the past 40 years \cite{anninga2011}. The strongest known prognostic factor for both event-free survival (i.e., time to local recurrence, metastatic disease, second malignancy, or death) and overall survival (i.e., time to death) in osteosarcoma is Histological Response (HRe) \cite{bishop2016}, i.e., the result of the histopathological examination to assess the improvement in microscopic tissue appearance following pre-operative chemotherapy. However, the impact of interventions in chemotherapy dosage and timing on patient survival remains unclear \cite{lancia2019method}.
In this study the primary research questions are: 

\begin{quote} \normalsize
	\textit{Does reduced chemotherapy dose intensity lead to an improvement in Event-Free Survival (EFS) of patients with osteosarcoma? Does this effect vary among subjects characterized by different histological responses?}
\end{quote}

Addressing these questions is very challenging, even with data from Randomized Clinical Trials (RCTs).
A first attempt was made in Lewis \textit{et al.}\cite{lewis2007}, where the authors investigated an Intention-To-Treat (ITT) landmark Cox model including as covariates the planned regimen, HRe, and their interaction. The ITT principle, widely applied in RCTs, measures the effect of \textit{assigning} patients to different regimens \cite{gupta2011,Smith2021}, disregarding post-randomization events, such as non-adherence or protocol deviations. 
However, the intensity of the assigned regimen often differs from the intensity of the received dose. 
Interventions and discontinuation in treatment administration are  common in actual clinical practice, due to the toxic side effects developed by patients over therapy \cite{souhami1997} which affect subsequent exposure by delaying the next cycle or reducing chemotherapy doses \cite{lancia2019,lancia2019novel}. Being at the same time risk factors for mortality and predictors of future exposure levels, toxicities are \textit{post-assignment confounders} for the effect of received dose intensity on patient’s survival.

To measure the discrepancies between \textit{assigned} (or planned) and \textit{received} (or actual) treatments in terms of both dose reduction and delays, the so-called Received Dose Intensity (RDI) indicator has been introduced \cite{RDIpaper}. 
Previous studies showed that there is a mismatch between planned and achieved chemotherapy-RDI in osteosarcoma \cite{lancia2019method,lancia2019novel}. Even patients assigned to the same regimen reported substantial variability in RDI at the end of treatment \cite{lancia2019method,lancia2019novel,spreaficoSMAP}.
To evaluate the impact of actually \textit{receiving} a treatment, per-protocol or as-treated analyses can be employed. The first focuses on participants who strictly adhered to the assigned protocol and excludes non-adherent data, while the second considers treatment actually received by patients, regardless of adherence to randomization \cite{Smith2021}. Nonetheless, both approaches compromise the balance between patient groups achieved through randomization, potentially introducing selection bias and confounding into the treatment effect estimate. In the presence of confounders, classical survival approaches \cite{cox1972,Therneau2010-cf,Kleinbaum2016-xr}  fail to estimate consistent causal effects. An alternative framework that emulates randomization, where confounders (e.g., toxicities) no longer predict treatment, is hence necessary. 

In clinical trials, interventions in treatment administration, as well as their underlying reasons, are typically well documented as required by protocols. This existing wealth of information has the potential to be repurposed for additional retrospective analyses beyond the scope of the original RCTs that generated the data, opening up new possibilities for further investigations.
More specifically, chemotherapy administration data can be employed to emulate another hypothetical RCT or Target Trial (TT) that explores new research questions on chemotherapy treatment outside the original scope. 
TT emulation has been introduced in Hern{\'a}n and Robins (2016)\cite{hernan2016} as a method for enabling the application of causal inference methods using observational data. A proper emulation requires a detailed specification of all the necessary protocol components (i.e., eligibility criteria, treatment strategies, treatment assignment, start and end of follow-up, outcomes, causal contrasts or estimands) and a data-analysis plan. This approach is particularly valuable for studying treatments or interventions where randomization is not possible or practical or is no longer present.

\subsection{Objectives}
In this article, a novel TT emulation based on RCT data of chemotherapy administration with interventions is proposed to estimate the effects of different received exposure strategies on EFS in patients with osteosarcoma aged 40 years or less at baseline. Three exposure strategies are defined and considered: 1) \textit{standard}, 2) \textit{reduced}, and 3) \textit{highly-reduced} RDI.
Data from two RCTs in osteosarcoma, namely, the European Osteosarcoma Intergroup (EOI) studies BO03 \cite{lewis2000} and BO06 \cite{lewis2007} (European Organisation for Research and Treatment of Cancer EORTC 80861 and 80931, respectively) are analysed. By considering patients originally assigned to the same chemotherapy regimen, i.e., the standard EOI treatment, it is shown how properly documented chemotherapy-administration data can be reused to address novel research questions.

A Marginal Structural Cox Model (Cox MSM) with effect modification estimated by using Inverse Probability of Treatment Weighting (IPTW) \cite{cibook2020} is employed to study a model similar to the ITT landmark Cox landmark model in Lewis \textit{et al.} (2007)\cite{lewis2007} in a causal setting. Specifically, the planned regimen in Lewis \textit{et al.} (2007)\cite{lewis2007} is replaced with our RDI-exposure strategies and their effect is supposed to vary based on the HRe (i.e., the effect modifier). 
IPTW is used to mimic randomization in the defined TT, where RDI-exposure is no longer confounded by toxicities or other confounders, so that a crude analysis suffices to estimate the effectiveness of RDI-reduction exposures on EFS in both HRe sub-groups. 
Conditional Average Treatment Effects (CATEs) are finally measured as the difference  between the Restricted Mean Survival Time (RMST) of \textit{reduced/highly-reduced} RDI strategy and the \textit{standard} one. A novel generalized bootstrap procedure \cite{efron1979,efrontib1994} utilizing unequal IPTW-based probability sampling \cite{DONALD2014,genbootstrap} and preserving the sizes of the sub-cohorts defined by different combinations of strategies and effect modifier levels is proposed to compute confidence intervals for CATEs.

The overall procedure hence requires (i) a proper definition of the RDI-exposure strategy, (ii) a tailor-made identification of all possible \textit{pre-assignement} and \textit{post-assignement} confounders, and (iii) a proper characterisation of the causal structure of the chemotherapy data through a Direct Acyclic Graph (DAG) \cite{greenland1999,cibook2020}.
Furthermore, since adjustments in treatment allocation are determined by the overall toxic burden of each patient, the different types and number of side effects must be adequately summarized and quantified. The new longitudinal Multiple Overall Toxicity (MOTox) score introduced in Spreafico \textit{et al.} (2021)\cite{spreaficoBMJOpen} is hence adapted to the data under study. This allows multiple toxicities to be included within the causal inference framework in a novel way.

The ultimate goal is to introduce an innovative and comprehensive RDI-based analysis of chemotherapy administration data with interventions. A tutorial-like explanations of the challenges inherent in this context is provided along with  novel problem-solving strategies. 
To the best of our knowledge, this study is the first to apply IPTW-based techniques to survival RCT data, aiming to mitigate the \textit{toxicity-treatment-adjustment} bias when estimating the effects of RDI reductions on EFS, while considering intrinsic personal responses to chemotherapy. Source code for the current study is available here: \url{https://github.com/mspreafico/TTEcausalRDI}.

\newpage
\section{Methods}\label{s:met}

\subsection{Data sources description: RCT data with interventions}
Data from control arms of European Osteosarcoma Intergroup (EOI) randomised clinical trials (RCT) BO03 and BO06 (EORTC 80861 and 80931, respectively) were analysed.
In both trials, control arms were characterized by the standard EOI treatment structured in 6 cycles of 3-weekly Cisplatin (CDDP) (100 $mg/m^2$) plus Doxorubicin (DOX) (75 $mg/m^2$), and compared to a different therapy regimen (i.e., variant of Rosen’s T10 regimen  in BO03 \cite{rosen1985} and a 2-weekly intensified version of CDDP+DOX in BO06 \cite{lewis2007}). Results about the primary analyses on BO03 and BO06 data can be found in Lewis \textit{et al.} (2000; 2007)\cite{lewis2000,lewis2007}.

\begin{figure}[h]	\centering{\includegraphics[width=1\textwidth]{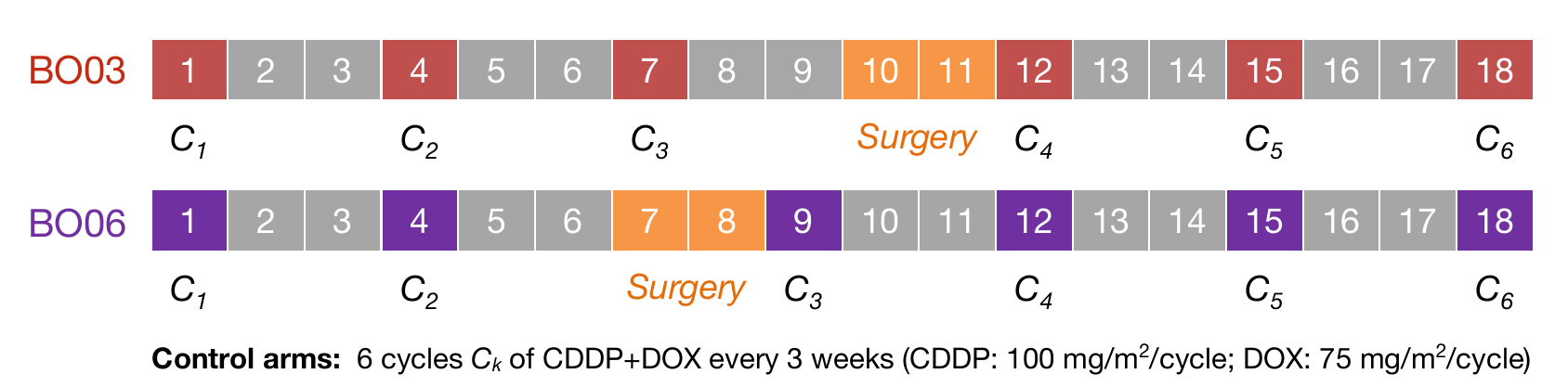}}
	\caption{Control arms design for BO03 and B006 randomised clinical trials, characterized by the standard European Osteosarcoma Intergroup treatment structured in 6 cycles of 3-weekly Cisplatin (CDDP) (100 $mg/m^2$) plus Doxorubicin (DOX) (75 $mg/m^2$).\label{fig:arms}}
\end{figure}

As the control arms design in Figure \ref{fig:arms} shows, in both trials chemotherapy was administered before and after surgical removal of the primary osteosarcoma.  At the end of the pre-operative treatment, with a nominal duration of 3 cycles in BO03 and 2 in BO06, the tumour was surgically resected, and the levels of tumour necrosis and Histological Response (HRe) evaluated.
Post-operative chemotherapy was intended to resume 2 weeks after surgery.

Along with patients baseline characteristics at randomization (age, gender, allocated chemotherapy regimen, site and location of the tumour), treatment-related variables (administered dose of chemotherapy, cycles timing, haematological parameters, chemotherapy-induced toxicity and histological response to pre-operative chemotherapy) were collected prospectively during therapy. 
These data provide insights into interventions made during therapy administration (i.e., cycle delays or dose reductions) and the associated toxicity reasons, that led the patient to deviate from the originally planned EOI chemotherapy regimen.

\subsubsection{Toxicity-driven interventions}
As it often occurs in clinical trials, therapy administration was complicated by the need for dynamic adjustments based on the patient's multi-systemic side effects (e.g., organ toxicity or myelosuppression) developed over time. Toxicities are a threat to patient’s life and must be controlled by either  allocating dose reductions/discontinuations or delaying the subsequent course \cite{lancia2019}.

Toxic side effects were recorded using the Common Terminology Criteria for Adverse Events Version 3 (CTCAE v3.0) \cite{ctcae3}, with grades ranging from 0 (none) to 4 (life-threatening) (see Appendix \ref{appA:ctcae} for further details). Toxicities were collected longitudinally in BO06 trial, whereas in BO03 only the highest CTCAE grade (i.e., the most severe) was recorded for each toxicity in both the pre-operative and post-operative periods. 
According to protocols, the following side effects were linked to specific dose reduction or delay rules: \textit{leucopenia}, \textit{thrombocytopenia}, \textit{oral mucositis}, \textit{ototoxicity}, \textit{cardiotoxicity} and \textit{neurotoxicity}. If different \textit{rule-specific} conditions co-existed and more than one dose reduction (or cumulative delays) applied, the lowest dose (or the highest delays) calculated was employed. 
According to expert knowledge, although not directly related to a specific adjustment rule, patient's \textit{generic} conditions of \textit{nausea/vomiting} and \textit{infections} was also taken into account during therapy. 

Interventions in treatment administration were hence determined as a combination of overall toxic burden related to both rule-specific and generic conditions. Toxicities impact patient's survival, leading to a complex post-assignment confounding mechanisms between received chemotherapy dose intensity and the outcome.

\subsubsection{Assessing interventions through Received Dose Intensity}
The so-called Received Dose Intensity (RDI) approach \cite{RDIpaper,lewis2000,lancia2019method} can be adopted
to evaluate both dose reductions/discontinuations, time-delays, and their impact in reducing the intensity over the whole therapy. This method summarizes information on treatment interventions by considering both received dose and actual timing. For each patient $i\in\{1,\dots,N\}$, RDI is defined as the ratio between standardized dose $\Delta_{i}$ and standardized time $\Gamma_{i}$, as follows:
\begin{equation}\label{eq:RDI}
	RDI_{i} = \frac{\Delta_i}{\Gamma_i}.
\end{equation}

Numerator in \eqref{eq:RDI} represents the \textit{standardized dose}, given by 
\begin{equation}\label{eq:Delta}
	\Delta_{i} = \frac{1}{2}	\left(	\Delta_{i}^{CDDP} +\Delta_{i}^{DOX} \right) = \frac{1}{12}	\left(	\sum_{j=1}^{6} \delta_{ij}^{CDDP} + \sum_{j=1}^{6} \delta_{ij}^{DOX} \right),
\end{equation}
where 6 is the total number of cycles in the EOI regimen, and $\delta_{ij}^{d}$ is the \textit{cycle-standardized received dose} defined as the ratio between  the actual dose $[mg/m^2]$ of drug $d \in \{\text{CDDP, DOX}\}$ assumed at cycle $j$ and the anticipated dose of drug $d$ (CDDP: 100 $mg/m^2$; DOX: 75 $mg/m^2$). Specifically, $\Delta_i<1$ indicates dose-reduced therapies, whereas $\Delta_i>1$ corresponds to dose-augmented therapies.

Denominator in \eqref{eq:RDI} represents the \textit{standardized time} given by
\begin{equation}\label{eq:Gamma}
	\Gamma_{i} = \frac{\text{actual treatment time}}{\text{anticipated treatment time}},
\end{equation}
where the \textit{actual treatment time} is the difference in days between the starting date of cycle 1 and the 3rd day after the start of cycle 6, and the \textit{anticipated treatment time} is $21 \times 5 + 14 + 3= 122$ days (i.e., 5 cycles lasting 21 days each, 14 days of surgery and 3 days after the start of cycle 6).
Specifically, $\Gamma_i>1$ indicates delayed therapies, whereas $\Gamma_i<1$ corresponds to compressed treatments.

In general, $\Delta_i \le 1$ and $\Gamma_i \ge 1$ due to dose reductions and delays, respectively; this implies  $RDI_i \le 1$. Based on expert knowledge, a RDI (in percentage) of at least 85\% is defined as \textit{standard} intensity level, from 85\% to 70\% is considered \textit{reduced}, whereas below 70\% is \textit{highly-reduced}.

\subsection{Study design}
To address the research questions, a target trial emulation approach is employed \cite{hernan2016}.
The protocol of the hypothetical TT and its emulation with chemotherapy administration data from BO03/BO06 RCTs are described in Table \ref{tab:TTE}. 
To be eligible, subjects have to be aged 40 years or less at baseline with a confirmed diagnosis of osteosarcoma. Further inclusion and exclusion criteria are applied to focus on the eligible cohorts of the original BO03/BO06 RCTs.
The TT compares three target strategies: 1) \textit{standard}, 2) \textit{reduced}, and 3) \textit{highly-reduced} RDI of the EOI control regimen given by 6 cycles of 3-weekly CDDP+DOX. The final aim is to study the effect, if any, of reductions in RDI (compared to standard) on EFS in subgroups of patients characterized by different HRe. 
Given that histopathological examination is evaluated after TT randomization, the statistical analysis has to be conducted utilizing a landmark approach\cite{landmark2007,landmarkbook,landmark2017} to appropriately incorporate HRe into the survival model. Specifically, an ITT landmark Cox model, with the landmark point at the time of surgery, is intended to serve as the survival model in the TT to estimate the effects of reduced exposures across HRe levels.

In the cohort selected from the BO03/BO06 data, randomization of target strategies is emulated by adjustment for confounding via IPTW. A pseudo-population is created by weighting each patient based on the inverse probability of observing a specific exposure allocation strategy given the confounders history. The pseudo-population mimics the randomized cohort of the TT and exhibits the following two properties:
\begin{enumerate}
	\item[i.] the pre-assignment and post-assignment history of pseudo-patients no longer predicts exposure to RDI-reductions in the next cycle;
	\item[ii.] the association between exposure and outcome is the same in both original and pseudo-population.
\end{enumerate}
Therefore, (heterogeneous) causal effects of different exposure strategies (across sub-groups defined by HRe) can be estimated by a crude analysis on the pseudo-population by using a Cox MSM with effect modifications.

\begin{table}
	\small
	\caption{Outline of the Target Trial protocol: specification and emulation using RCT data with interventions.}\label{tab:TTE}
	\begin{tabular*}{\textwidth}{@{\extracolsep\fill}p{2.1cm}p{5.6cm}p{5.6cm}}
		\hline
		& \textbf{Target trial} & \textbf{RCT data with interventions} \\ 
		\multirow{-2}{2.1cm}{\textbf{Protocol components}} & Specification & Emulation \textit{[Required data]}\\
		\hline
		Aim & To estimate the effect, if any, of reductions in received dose intensity on event-free survival across levels of histological response in patients with osteosarcoma aged 40 years or less. & Same as for TT specification. \\
		\hline
		Eligibility \newline criteria & Patients aged 40 years or less with a histologically confirmed diagnosis of high-grade osteosarcoma in an extremity long bone.\footnotemark[1] Patients need to commence chemotherapy within 28 days after biopsy, with normal leukocyte ($\ge 3.5\times10^9/L$) and platelet ($\ge 100\times10^9/L$) counts.  & Same as for TT specification \newline \textit{[Age, diagnosis type, start date of therapy, leukocyte count, platelet count]}\\ 
		\hline
		Treatment  \newline strategies & Chemotherapy dose intensity regimens: \textit{standard}, \textit{reduced}, and \textit{highly-reduced} EOI treatment.\footnotemark[2] & Same as for TT specification \newline \textit{[Received dose, cycle timing]} \\ 
		\hline
		Treatment  \newline  assignments & Eligible persons will be randomly assigned to one strategy and will be aware of which strategy they were assigned to. & 
		Eligible persons assigned to the strategy based on observed dose-intensity behavior at the end of the therapy.\\ 
		\hline
		Outcomes & Death, local recurrence, evidence of new or progressive metastatic disease, second malignancy, or a combination of those events.& Same as for TT specification \newline \textit{[Date of death, local recurrence, or other malignancies]}  \\ 
		\hline
		Follow-up & Start: treatment assignment \newline End: death, local recurrence, evidence of new/progressive metastatic disease or second malignancy, or censoring. & Same as for TT specification, except start is the end of therapy \newline \textit{[Date of loss to follow-up or censoring]}  \\ 
		\hline
		Causal \newline estimand & Intention-to-treat conditional effects (effect of being assigned to a reduced treatment across subgroups) \newline Per-protocol conditional effect (effect of receiving a reduced treatment as indicated in the protocol across subgroups). & Observational analogue of per-protocol conditional effect across subgroup.\\ 
		\hline
		Statistical analysis & Intention-to-treat analysis via landmark Cox model \cite{landmark2007,landmarkbook,landmark2017} to estimate effects of reduced exposures across levels of histological response.\footnotemark[3] \newline
		Conditional average effect measured as contrast of restricted mean survival times over follow-up. & Per-protocol analysis analogous to TT specification. \newline Randomization will be emulated via adjustment for pre-assignment and post-assignment confounders by inverse probability of treatment weighting. \newline \textit{[Pre-assignement confounders, post-assignment confounders, histological response]}  \\ 
		\hline
	\end{tabular*}
\begin{tablenotes}
	\scriptsize
	\item $^1${Ineligible patients: subjects with paraosteal, periosteal, Paget-related, or radiation-induced osteosarcoma; patients with prior malignancy, any chemotherapy before trial entry, reduced glomerular filtration rate ($<60$ mL/min/1.73 m$^2$), cardiac dysfunction, or raised bilirubin.}
	\item $^2${Standard EOI treatment: 6 cycles of 3-weekly CDDP (100 $mg/m^2$) plus DOX (75 $mg/m^2$).}
	\item $^3${Samilarly to the ITT landmark Cox model for regimen effect stratified by histological response in Lewis \textit{et al.}\cite{lewis2007}.}
	\end{tablenotes}
\end{table}

\subsection{Causal inference framework}
To address the research questions at hand, it is imperative to appropriately emulate the target causal inference framework and develop a suitable data analysis plan. This requires both clinical expertise in the treatment of osteosarcoma and statistical knowledge in variable definition and mathematical modeling. 
Causal analysis involving effect modification focuses on investigating the causal relationship between exposure and outcome across various levels of another factor that impacts this connection, and it requires adjustment for exposure-outcome confounders. The components of our causal framework hence include exposure, outcome, confounders, and effect modifier, as defined in the following sections. The causal structure is finally represented through a Directed Acyclic Graph (DAG) \cite{greenland1999,cibook2020}. This process requires special attention to the identifiability assumptions \cite{cibook2020, ravani2017} of \textit{consistency}, \textit{no unmeasured confounding}, and \textit{positivity}, discussed in details in Appendix \ref{appB:CIassumptions}.

\subsubsection{Outcome}
The endpoint of this study is EFS, defined as time from the end of therapy until the first event (local recurrence, evidence of new or progressive metastatic disease, second malignancy, death, or a combination of those events) or censoring at last contact.
Let $T_i$ = $\min(T_i^*,C_i)$ be the observed EFS time, where $T_i^*$ is the true event time, and $C_i$ is the censoring time (i.e., the time from the end of the therapy until the last visit). Let $D_i = I(T_i^* \le C_i)$ be the event indicator (1 when $T_i^* \le C_i$, and 0 otherwise).
The EFS outcome for patient $i \in \{1,...,N\}$ is denoted by the pair $(T_i,D_i)$.

\subsubsection{Exposure}
The exposure strategies related to RDI values are now defined based on expert knowledge. A RDI percentage of 85\% or more is considered a \textit{standard} intensity level, as reductions up to 15\% are classified as negligible. This standard level can be compared to reductions ranging from 15\% to 30\% (\textit{reduced} intensity) and reductions above 30\% (\textit{highly-reduced} intensity). Consequently, covariate $A_i$ for RDI-exposure is defined as a three-level categorical variable, as follows:
\begin{equation}\label{eq:exp:rdi}
	A_i = \begin{cases}
		0 \quad \text{if $RDI_i\ge0.85$}\\
		1 \quad \text{if $0.70 \le RDI_i<0.85$}\\
		2 \quad \text{if $RDI_i<0.70$}
	\end{cases}
\end{equation}
that is,  $A_i =0$ is equivalent to a \textit{“standard”} RDI,  $A_i =1$ to a \textit{“reduced”} RDI, and  $A_i =2$ to a \textit{“highly-reduced”} RDI. Accordingly, the three possible treatment/exposure strategies are denoted by $a \in \{0, 1, 2\}$. Based on expert knowledge, these strategies are well-defined to ensure the \textit{consistency} assumption (see Appendix \ref{appB:CIassumptions}).

\subsubsection{Effect modifier}
Effect modification focuses on subgroup-specific causal effects of a single type of exposure \cite{cibook2020,bours2021}. In general, a modifying variable $V$ should be included into the analysis under two conditions \cite{cibook2020}: (i) when the investigators believe that $V$ could potentially act as an effect modifier; (ii) when the investigators are more interested in understanding the causal effect of exposure within the groups defined by covariate $V$ rather than examining it across the entire population. In the application considered here, variable $V_i$ is the binary covariate representing the HRe of subject $i$, as defined in the original RCTs:
\begin{equation}\label{eq:effmod:hre}
	V_i = \begin{cases}
		0 \quad \text{if tumour necrosis$_i<90\%$}\\
		1 \quad \text{if tumour necrosis$_i\ge90\%$}\\
	\end{cases}
\end{equation}
that is, $V_i=1$ for patients with a \textit{“good”} HRe, i.e., Good Responders (GRs), while $V_i=0$ denotes patients with a \textit{“poor”} HRe, i.e., Poor Responders (PRs).

\subsubsection{Confounders}
To draw valid conclusions about the causal exposure effect, the set of confounders of the exposure-outcome relationship under study need to be considered in the analysis. 
According to experts knowledge and protocol guidelines, the following pre-assignment and post-assignment characteristics, denoted by vector $\boldsymbol{L}_i$, satisfy the hypothesis of \textit{no unmeasured confounding} (see Appendix \ref{appB:CIassumptions}).\\

\textbf{\textit{Pre-assignment confounders.}} Due to their potential influence on drug metabolism and increased toxicity risk, \texttt{age} group, as defined in Collins \textit{et al.} (2013)\cite{collins2013}  (\textit{child}: 0–12/0-11 years for males/females; \textit{adolescent}: 13–17/12–16 years for males/females; \textit{adult}: 18/17 or older for males/females), as well as \texttt{gender} (\textit{female}; \textit{male}) serve as pre-assignment confounders. While the \texttt{trial} number (BO03; BO06) does not serve as a significant risk factor for failures (p-value of log-rank test for Kaplan-Meier estimators stratified by trial is 0.967 -- see Table \ref{tab:descriptive}), it can still be considered a pre-assignment confounder, as it reflect the different number of pre-operative cycles (see Figure \ref{fig:arms}) and independently predicts dose intensity (p-value of chi-squared test for the association between RDI-exposure and trial cohorts is $<0.001$ -- see Table \ref{tab:descriptive}). \\

\textbf{\textit{Post-assignment confounders.}} 
Conditioning chemotherapy administration over treatment, \textit{rule-specific} and \textit{generic} toxicities are post-assignment confounding factors.
To properly address toxicities as confounding covariates, it is essential to accurately quantify and summarize the pre- and post-operative overall toxic burden arising from individual CTCAE side effects.
This is achieved by utilizing the new longitudinal Multiple Overall Toxicity (MOTox) score \cite{spreaficoBMJOpen}. The MOTox score incorporates three significant components of adverse events: (i) multiple lower-grade chronic toxicities (which may affect the patient's quality of life); (ii) substantial level in a specific toxicity (potentially causing severe and permanent consequences for the patient); (iii) time dependency.

Since toxicity data over cycles were not recorded for the BO03 trial, MOTox computation is based on pre- and post-operative periods, by considering the highest CTCAE grade recorded for each toxicity during  pre/post-operative cycles.

Let $\mathcal{M}_{rule} = \{$\textit{leucopenia}, \textit{thrombocytopenia}, \textit{oral mucositis}, \textit{ototoxicity}, \textit{cardiotoxicity}, \textit{neurotoxicity}$\}$ and $\mathcal{M}_{gen} = \{$\textit{nausea}, \textit{infection}$\}$ be the two disjoint sets of toxicities related to \textit{rule-specific} and \textit{generic} toxicities, respectively. Denote by $k \in \{pre,post\}$ the pre/post-operative time-period. For each patient $i$, let $tox_{ijk}^{m}$ (with value from 0 to 4) be the most severe CTCAE grade of the $m$-th toxicity of type $j \in \{rule,gen\}$ (with $m=1,...,|\mathcal{M}_{j}|$) measured during period $k$. The \texttt{MOTox} score related to set $\mathcal{M}_{j}$ for the $i$-th patient during period $k$ is defined as follows:
\begin{equation}\label{eq:motox}
	MOTox_{ijk} = \frac{1}{|\mathcal{M}_j|} \sum_{m=1}^{|\mathcal{M}_j|} tox_{ijk}^{m} + \max_{m=1,...,|\mathcal{M}_j|}{\left(tox_{ijk}^{m}\right)}.
\end{equation}
Specifically, four different MOTox scores can be computed for each subject.

By adopting this approach rather than relying on individual CTCAE grades for diverse toxicities, problems associated with dealing with a vast number of potential confounder combinations are mitigated. This ensures increased feasibility in the analysis. When considering individual grades for each toxicity, the number of possible confounders combinations would be too high leading to a violation of positivity. Additionally, this approach alignes with clinical practices, where treatment adaptation occurs based on the patient's overall toxic burden due to the presence of multiple toxicities.

\subsubsection{Directed Acyclic Graph (DAG)}
\begin{figure}[t]%
	\centering
	\includegraphics[width=1\textwidth]{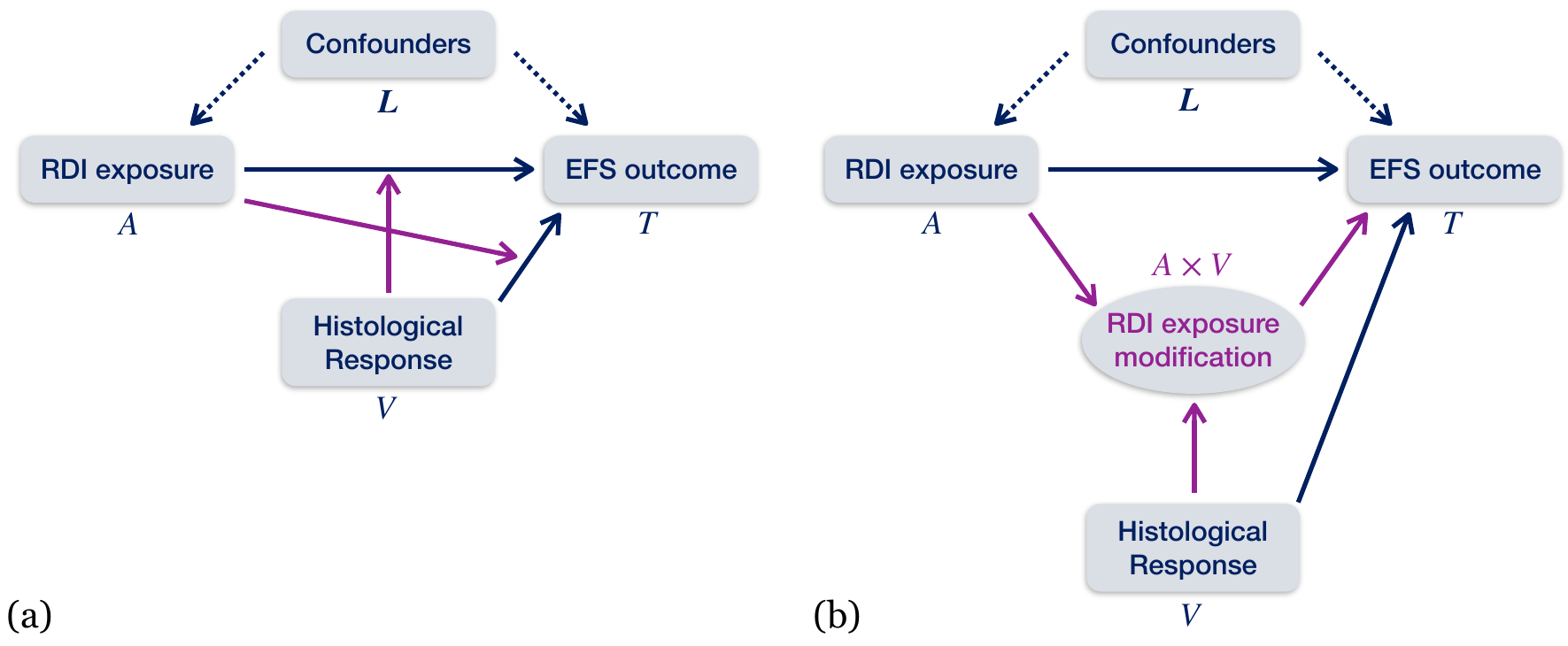}
	\caption{Directed Acyclic Graph (DAG) that represents the causal relationships between EFS outcome ($T$), RDI-exposure ($A$), pre-/post-assignment confounders ($\boldsymbol{L}$), and HRe as effect modifier ($V$). The exposure-effect modification pathway $A\times V \rightarrow T$ (in purple) is depicted in DAG (a) using the ``arrow-on-arrow" representation proposed by Weinberg (2007)\cite{weinberg2007}, whereas in DAG (b) including the additional node with both $A$ and $V$ as parents, as suggested by Attia \textit{et al.} (2022)\cite{attia2022}.}\label{fig:dag}
\end{figure}

Figure \ref{fig:dag} presents two alternative visualizations of the causal structure involving RDI-exposure ($A$), EFS outcome ($T$), pre- and post-assignment confounders ($\boldsymbol{L}$), and HRe as a modifying variable ($V$).
In both cases, blue solid arrows indicate that both exposure $A$ and the effect modifier $V$ directly influence the outcome $T$, while dashed blue arrows represent the confounding relationship between $A$ and $T$.
The purple arrows represent the influence of exposure-effect modification $A\times V$ on $T$, but there is no unanimous consensus on how to graphically represent $A\times V \rightarrow T$. In this regard, DAG (a) utilizes the crossing ``arrow-on-arrow" representation provided by Weinberg (2007)\cite{weinberg2007}, while DAG (b) includes an additional node with both $A$ and $V$ as parents, as proposed in Attia \textit{et al.} (2022)\cite{attia2022}.

The causal structure relies upon the hypothesis that there is no path between HRe and RDI-exposure, i.e., $A \centernot{\leftrightarrow} V$. This assumption is motivated by the following reasons.
\begin{enumerate}
	\item HRe is the result of the histopathological examination after pre-operative chemotherapy. This means that RDI computed as in Equation \eqref{eq:RDI} at the end of treatment (i.e., after both the pre- and the post-operative periods) could not affect HRe. This was confirmed by the absence of evidence indicating an association between the final RDI and HRe (see Figure \ref{fig:RDI_delta}), as reported in Lewis \textit{et al.} (2000)\cite{lewis2000} as well. Therefore $A \centernot{\rightarrow} V$.
	\item In the original BO03/BO06 RCTs, HRe was not known until several weeks since chemotherapy is resumed after surgery. This means that HRe result could have influenced the decision to modify therapy only in the last cycles. However, the original RCT protocols did not provide for treatment interventions based on HRe. As clinicians are generally committed to adhering to the planned treatment without being influenced by factors not foreseen in the protocol, very few protocol violations are expected in a RCT. Therefore, $V \centernot{\rightarrow} A$.
\end{enumerate}

\subsection{Statistical Analysis}
Once the causal inference framework has been defined, statistical analysis can be performed. This requires careful consideration about the identifiability assumptions \cite{cibook2020,ravani2017} related to \textit{positivity} and \textit{absence of model misspecification} (see Appendix \ref{appB:CIassumptions}).
Building upon the ITT landmark Cox model examined in Lewis \textit{et al.} (2007)\cite{lewis2007}, the idea is to assess subgroup-specific causal effects of different RDI-exposure stategies on EFS-time using a Cox MSM with effect modification. In the ITT model from Lewis \textit{et al.} (2007)\cite{lewis2007}, the analysis incorporated the intended treatment, HRe, and their interaction to investigate the effect of assigned regimens stratified by HRe. In the Cox MSM proposed here, the binary variable representing intended treatment is replaced by two dummy variable representing \textit{reduced} and \textit{highly-reduced} RDI strategies, and their effects are assumed to vary based on the effect modifier (HRe).

\subsubsection{Marginal structural Cox model with effect modification}
Cox MSMs are a class of causal models that focus on \textit{counterfactual} time-to-event variables \cite{hernan2000epi,robins2000,cibook2020}. These variables represent the time at which an event would have been observed had a patient been administered a specific exposure level $a$, which might differ from the actual treatment received.
In our context, the \textit{counterfactual EFS time} that would be observed in a subject under exposure $a \in \{0,1,2\}$ is denoted by $T^{a}$.

The Cox-type marginal structural hazard function for counterfactual EFS time under RDI-exposure $a\in \{0:\textit{standard}; \, 1:\textit{reduced};\, 2:\textit{highly-reduced}\}$ with effect modification given by HRe variable $V \in \{0:\textit{poor};\,  1:\textit{good}\}$ is defined as follows:
\begin{equation}\label{eq:MSCM1}
	h_{T^{a}}(t|V) = h_0(t) \exp\left\{\beta_1 \mathbbm{1}_{(a=1)} + \beta_2 \mathbbm{1}_{(a=2)} + \beta_3 \mathbbm{1}_{(a=1)}V  + \beta_4 \mathbbm{1}_{(a=2)} V + \beta_5 V 
	\right\}.
\end{equation}
Additive effect modification is present for a \textit{reduced} RDI if $\beta_3\neq0$ or for a \textit{highly-reduced} RDI  if $\beta_4\neq0$. 

Evidence for effect modification aids in identifying groups of individuals with specific inherent characteristics which make them better responsive to treatment, while in others, treatment may be less effective, ineffective, or even harmful \cite{bours2021}.

\subsubsection{Inverse Probability of Treatment Weighting (IPTW)}
To estimate the causal parameters $\boldsymbol{\beta}$ of the Cox MSM defined in \eqref{eq:MSCM1}, a weighted Cox model \cite{Binder1992,Lin2000} can be fitted to the pseudo-population obtained through IPTW, as follows:
\begin{equation}\label{eq:cox1}
	h^{SW_i }_{T_i}\left(t | A_i, V_i\right) = h_0(t) \exp\left\{\theta_1 \mathbbm{1}_{(A_i=1)} + \theta_2 \mathbbm{1}_{( A_i=2)} + \theta_3  \mathbbm{1}_{(A_i=1)}V_i  + \theta_4 \mathbbm{1}_{(A_i=2)} V_i + \theta_5 V_i \right\}
\end{equation}
with subject-specific stabilized weights  given by 
\begin{equation}\label{eq:sw1}
	SW_i= \frac{ P\left(A_i | V_i \right) }{ P\left(A_i \big|  \boldsymbol{L}_i, V_i\right)}.
\end{equation}
The numerator in \eqref{eq:sw1} represents the probability that a subject $i$ received exposure $A_i$ given their HRe $V_i$. Including the effect modifier in the numerator generally results in narrower confidence intervals around the effect estimates \cite{cibook2020}. 
The denominator is the probability that the subject received exposure $A_i$ given HRe and confounders. In this case, the effect modifier is included to enhance the efficiency of the MSM parameter estimation process, as recommended in Hern\'{a}n and Robins (2020)\cite{cibook2020}.
Both numerator and denominator are modelled by employing multinomial logistic regression models. 

Under causal inference assumptions, association is causation in the pseudo-population and the estimates of the associational parameters $\boldsymbol{\theta}$ are consistent for the causal parameters $\boldsymbol{\beta}$. Nonetheless, a note of caution is required  in applying this methodology to the chemotherapy data. Different model specifications in terms of confounding covariate features must be compared to satisfy the final assumptions of \textit{positivity} and \textit{no misspecification of the weight-generating models} (see Appendix \ref{appB:CIassumptions}) and guarantee an unbiased estimation of the results. Specifically, a mean weight value that significantly deviates from one or the presence of extreme values in the distribution of the stabilized weights can signal potential issues related to positivity violation or model misspecification \cite{cole2009epi}. In addition, graphical methods can be employed to check covariate balance between the exposure groups in the weighted samples \cite{austinstuart2015}.

\subsubsection{Conditional Average Treatment Effects (CATEs)}
The Restricted Mean Survival Time (RMST) \cite{Royston2011,Uno2014} is employed as measure of treatment effect. More precisely, the RMST at time $t$ under strategy $a\in \{0,1,2\}$ for individuals in sub-group $v\in \{0,1\}$ is the expected conditional time-to-event defined as follows:
\begin{equation}
	\mu_a(t;v) = \mathbbm{E}[\text{min}\{T^a,t\} | V=v] = \int_0^{t} S^a(s| V=v)ds.
\end{equation}
This corresponds to the area under the counterfactual survival curve given the effect modifier $V$ truncated at time $t$.

The Conditional Average Treatment Effect (CATE) at time $t$, or the ``benefit" in each HRe sub-group, is measured as the contrast between the RMSTs of an RDI-reduction intervention ($a={1,2}$) and the \textit{standard} strategy, as follows:
\begin{equation}\label{eq:cate}
	\tau_{a}(t;v) = \mu_{a}(t;v) - \mu_{0}(t;v) \qquad a\in \{1,2\}, v\in \{0,1\}. 
\end{equation}
CATE is hence an estimate of the average months gained (if $>0$) or lost (if $<0$) at time $t$ by employing RDI-reduction strategy $a\in \{1,2\}$ in sub-group $V=v$. 

\subsubsection{IPTW-based bootstrap procedure for estimating confidence intervals for CATEs}
To construct 95\% point-wise Confidence Intervals (CIs) for each CATE, a generalized bootstrap procedure is proposed. This novel sampling procedure differs from typical random sampling by (i) separately considering the sub-cohorts defined by different combinations of strategies and effect modifier levels, (ii) utilizing unequal probability sampling \cite{DONALD2014,genbootstrap} based on estimated IPTW stabilized weights, and (iii) sampling (with repetitions) from each sub-cohort while maintaining sub-sample sizes. At each iteration the generalized bootstrap sample is generated as the union of the various bootstrap sub-samples. The steps are detailed as follows.

\begin{enumerate}
	\item Determine the set of possible sub-cohorts: $$\mathcal{G} = \left\{(a,v):\, a=0,1,2;\, v=0,1\right\}$$
	
	\item Assign the subjects to the sub-cohorts $g\in\mathcal{G}$:
	$$\mathcal{D}_g = \left\{i \in \{1,\dots,N\}: (A_i, V_i)=g\right\} \quad  \text{ with sample size } n_g = |\mathcal{D}_g|.$$
	
	\item For each sub-cohort $g\in \mathcal{G}$, compute the sampling probability of each subject $j\in \mathcal{D}_{g}$ as a transformation of their stabilized weight from IPTW-Equation \eqref{eq:sw1} as follows:
	$$p_{gj} = \frac{sw_j}{\sum_{k=1}^{n_{g}} sw_k}.$$
	These unequal sampling probabilities represent the normalized IPTW stabilized weights within the sub-cohort $g$ in such a way that $\sum_{j\in\mathcal{D}_{g}} p_{gj} = 1$.
	
	\item At each bootstrap iteration $b=1,\dots, B$ (with $B=1000$):
	\begin{enumerate}
		\item obtain the sub-samples $\mathcal{D}^b_{g}$ with $n_g$ subjects sampled with repetitions from $\mathcal{D}_{g}$, where each subject $j$ has probability $p_{gj}$ to be selected;
		\item  combine the sub-samples $\mathcal{D}_{g}^b$ into the generalized bootstrap sample $ \mathcal{D}^b$: $$ \mathcal{D}^b = \bigcup_{g \in \mathcal{G}} \mathcal{D}_g^b \quad \text{ where } \quad |\mathcal{D}^b| = \sum_{g \in \mathcal{G}} n_g = N;$$
		\item Estimate the CATEs $\tau_{a}^b(t;v)$ over time $t$ in \eqref{eq:cate}  on the generalized bootstrap sample $\mathcal{D}^b$.
	\end{enumerate}
	
	\item For each RDI-reduction strategy $a\in\{1,2\}$, effect modifier stratum $v\in\{0,1\}$ and time-point $t$, the estimates $\hat\tau_{a}^b(t;v)$ are ordered from smallest to largest. The resulting 2.5th and 97.5th percentiles are selected to define the bounds of the 95\% bootstrap CI \cite{efron1979,efrontib1994}. 
\end{enumerate}

\section{Results}\label{s:res}
Statistical analyses were performed in the {\tt{R}}-software environment \cite{R}, in particular using {\tt{ipw}} \cite{ipw} and {\tt{survival}} \cite{survival} packages. Source code for the current study is available here: \url{https://github.com/mspreafico/TTEcausalRDI}.

\subsection{Study cohort}
In total 444 eligible patients were enrolled in the control arms of BO03 (199) and BO06 (245). In this sample, 106 (23.9\%) patients were excluded due to missing HRe. Among the remaining 338 patients, 58 subjects stopped the chemotherapy treatment or did not undergo surgery, while 4 completed the treatment but experienced an event during its administration. The final cohort of 276 patients (114 from BO03 and 162 from BO06, respectively) included in the per-protocol analyses (62.2\% of the initial sample) is shown in the consort diagram in Appendix \ref{appC:cohort}.

\subsection{Descriptives}

Patient characteristics over the entire cohort and by trial are shown in Table \ref{tab:descriptive}. 
Overall, the median RDI value was 0.759 (IQR=[0.649; 0.857]), with minimum and maximum values of 0.376 and 1.121. This corresponded to a total of 75 patients (27.2\%) with \textit{standard} RDI, 111 (40.2\%) with \textit{reduced} RDI, and 90 (32.6\%) with \textit{highly-reduced} RDI. Median EFS time computed using the reverse Kaplan-Meier method \cite{schemper1996} was 89.59 months (IQR = [50.33; 146.30]) and 152 patients (55.1\%) experienced an event after the end of the therapy.
\textit{Generic} MOTox scores were high: pre/post-operative median MOTox values were equal to 4.5; this means that in median patients experienced at least one generic side effect of CTCAE-grade 3 (i.e., severe or medically significant). This is not surprising because  nausea is the most common chemotherapy-induced adverse event.
\textit{Rule-specific} MOTox resulted higher in the post-operative period than in the pre-surgery one. This indicated that toxicity levels accumulate over time resulting in a more severe overall toxic burden in the second phase of treatment. A total of 94 patients (34.1\%) experienced \textit{good} HRe after surgical resection.

\begin{table}
	\small
	\caption{Patients and trial characteristics.}\label{tab:descriptive}
	\begin{tabular*}{\textwidth}{@{\extracolsep\fill}lcccr}
		\hline
		& \textbf{All}  & \textbf{BO03}  & \textbf{BO06} & \\ 
		\textbf{Patients} &  276  & 114 (41.3\%) & 162 (58.7\%) & \textbf{p-value}\footnotemark[1]\\
		\hline
		
		\textbf{Age}\footnotemark[2]  &&&& 0.259\\
		~ \textit{child} &  76 (27.5\%) & 26 (22.8\%) & 50 (30.9\%) & \\
		~ \textit{adolescent} &  117 (42.4\%) & 49 (43.0\%) & 68 (42.0\%) &\\
		~ \textit{adult}  &  83 (30.1\%) & 39 (34.2\%) & 44 (27.1\%) &\\
		
		\textbf{Gender} &&&& 0.703\\
		~ \textit{Female} &  109 (39.5\%) & 43 (37.7\%) & 66 (40.7\%) & \\
		~ \textit{Male} &  167 (60.5\%) & 71 (62.3\%) & 96 (59.3\%) &\\
		
		\textbf{Rule-specific MOTox}\footnotemark[3] &&&& \\
		~ \textit{\textbf{Pre-operative}} &&&& 0.009\\
		~\quad Median & 3.667 & 3.883  & 3.500  & \\
		~\quad IQR & $[2.500; 4.500]$ & $[2.667; 4.792]$ & $[2.333; 4.167]$ & \\
		~\quad Min/Max & 0/6.167 & 0/6.167 & 0/5.833 &\\
		~ \textit{\textbf{Post-operative}} &&&& $<0.001$\\
		~\quad Median & 4.333 & 3.833 & 5.000 & \\
		~\quad IQR & $[3.667; 5.333]$ & $[2.500; 4.958]$ & $[4.000; 5.500]$ & \\
		~\quad Min/Max & 0/6.833 & 0/6.167 & 0/6.833 &\\

		\textbf{Generic MOTox}\footnotemark[4] &&&& \\
		~ \textit{\textbf{Pre-operative}} &&&&  0.017\\
		~\quad Median & 4.500  & 4.500   & 4.000   & \\
		~\quad IQR &  $[3.000; 5.500]$ &  $[3.500; 5.375]$ &  $[3.000; 5.500]$ & \\
		~\quad Min/Max & 0/8 & 0/8 & 0/7.500 &\\
		~ \textit{\textbf{Post-operative}} &&&& 0.021\\
		~\quad Median & 4.500 & 4.500  & 4.000  & \\
		~\quad IQR &  $[3.000; 5.500]$ &  $[3.500; 5.375]$ &  $[3.000; 5.500]$ & \\
		~\quad Min/Max & 0/8 & 0/7.500 & 0/8 &\\
		
		\hline
		\textbf{Histological Response} &&&& 0.732\\
		~ \textit{poor} &  182 (65.9\%) & 77 (67.5\%) & 105 (64.8\%) & \\
		~ \textit{good} &  94 (34.1\%) & 37 (32.5\%) & 57 (35.2\%) &\\
		
		\hline
		\textbf{RDI} &&&& $<0.001$\\
		~\quad Median & 0.759 & 0.692  & 0.805  & \\
		~\quad IQR &  $[0.649; 0.857]$ &  $[0.589; 0.779]$ &  $[0.762; 0.899]$ & \\
		~\quad Min/Max & 0.376/1.121 & 0.376/1.028 & 0.424/1.121 &\\
		
		\textbf{RDI exposure} &&&& $<0.001$\\
		~ \textit{standard} &  75 (27.2\%) & 10 (8.8\%) & 65 (40.1\%) & \\
		~ \textit{reduced} &  111 (40.2\%) & 46 (40.3\%) & 65 (40.1\%) &\\
		~ \textit{highly-reduced} &  90 (32.6\%) & 58 (50.9\%) & 32 (19.8\%) &\\
		
		\hline
		\textbf{EFS status} &&&& 0.673\\
		~ \textit{censored} &  124 (44.9\%) & 49 (43.0\%) & 75 (46.3\%) & \\
		~ \textit{with event} &  152 (55.1\%) & 65 (57.0\%) & 87 (53.7\%) &\\
		
		\textbf{EFS time} [months] &&&& 0.962\\
		~\quad Median\footnotemark[5] [IQR] & 89.59 $[50.33;146.30]$ &&&\\
		
		\hline
	\end{tabular*}
\begin{tablenotes}
	\scriptsize
	\item $^1$Categorical variables: p-value of chi-squared test for association with the trial. Continuous variables: p-value of two-sided Mann-Whitney U test for the variable distribution in BO03 vs BO06 cohort. EFS time: p-value of log-rank test for Kaplan-Meier estimators stratified by trial.
	\item$^2${Age groups were defined according to  Collins \textit{et al.} (2013)\cite{collins2013}: \textit{child} (male: 0–12 years; female: 0–11 years), \textit{adolescent} (male: 13–17 years; female: 12–16 years) and \textit{adult} (male: 18 or older; female: age 17 years or older).}
	\item$^3${Pre-/post-operative MOTox scores computed using Equation \eqref{eq:motox} based on  \textit{rule-specific} conditions:  $\mathcal{M}_{rule} = \{$\textit{leucopenia}, \textit{thrombocytopenia}, \textit{oral mucositis}, \textit{ototoxicity}, \textit{cardiotoxicity}, \textit{neurotoxicity}$\}$.}
	\item$^4${Pre-/post-operative MOTox scores computed using Equation \eqref{eq:motox} based on  \textit{generic} conditions: $\mathcal{M}_{gen} = \{$\textit{nausea}, \textit{infection}$\}$.}
	\item$^5${Median EFS time was computed using the reverse Kaplan-Meier method \cite{schemper1996}.}
\end{tablenotes}		
\end{table}

\begin{figure}[t]
	\centering{\includegraphics[width=1\textwidth]{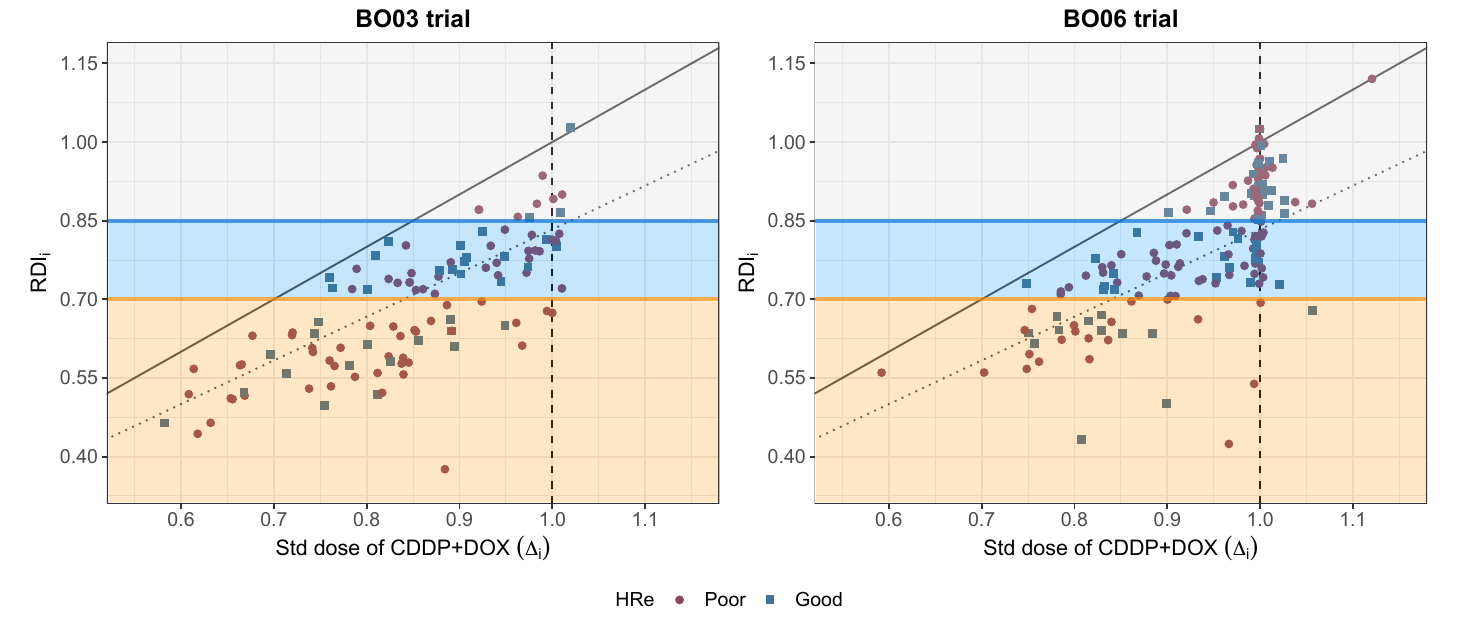}}
	\caption{Scatter plots of $RDI_i$ against the standardized dose $\Delta_i$ of CDDP+DOX conditional on trial (left panel: \textit{BO03}; right panel: \textit{BO06}) and HRe (purple points: \textit{poor}; blue squares: \textit{good}).\label{fig:RDI_delta}}
\end{figure}

Figure \ref{fig:RDI_delta} shows a scatter plot of $RDI_i$ against the standardized dose $\Delta_i$ of CDDP+DOX for each trial (left panel: \textit{BO03}; right panel: \textit{BO06}) and HRe (circles: \textit{poor}; squares: \textit{good}). Points to the left of the black dashed vertical line, where $\Delta_i<1$, represent patients who received dose-reduced therapies. The black diagonal solid line satisfies equation $RDI_i = \Delta_i$, dividing the group of patients with standardized time $\Gamma_i>1$ (delayed therapy, below the line) from the group of patients with $\Gamma_i<1$ (anticipated therapy, above the line). The black diagonal dotted line satisfies equation $RDI_i = \Delta_i/1.2$, dividing the group of patients with therapy delayed by more than 20\% of anticipated time (below the dotted line) from the group of patients with therapy delayed by less than 20\% of anticipated time (between solid and dotted black lines). 
Solid horizontal lines vertically divide patients with a \textit{standard} RDI-exposure (grey area above the blue line) from those with \textit{reduced} (blue area between the blue and orange lines) and \textit{highly-reduced} exposure (orange area below the orange line).
This figure shows lack of a clear association between HRe and RDI-exposure, as confirmed by the chi-squared test (p-value = 0.614). 

\subsection{IPTW diagnostics}
Multinomial logistic regressions were used to model both numerators and denominators of stabilized weights $SW_i$ in \eqref{eq:sw1}. Five different IPTW specifications in terms of the confounding covariates included in the denominators (Table \ref{tab:IPTW}) were investigated  to determine whether and which models best satisfied \textit{positivity} and \textit{no misspecification}. See Appendix \ref{appD:IPTW} for further details.

\begin{table}[h!]
	\small
	\caption{Inverse Probability of Treatment Weighting (IPTW) diagnostics based on summaries of stabilized weights $SW_i$ by different specifications of multinomial logistic regressions for the denominator $\Pr\left(A_i \big|  \boldsymbol{L}_i\right)$.}\label{tab:IPTW}
	\begin{tabular*}{\textwidth}{@{\extracolsep\fill}cp{7.5cm}cc}
		\hline%
		\multicolumn{2}{@{}c@{}}{\textbf{IPTW Specification}} & \multicolumn{2}{@{}c@{}}{\textbf{Stabilized weights: $SW_i$}} \\ 
		\textit{Method} & \textit{Description}\footnotemark[1] & \textit{Mean (s.d.)}  & \textit{Min/Max}  \\
		\hline
		IPTW 1 & Categorical/binary confounders and binary effect modifier as main effect only; each continuous MOTox score linearly related to the log-odds. & 0.988 (0.663) & 0.311/5.154 \\ \hline
		IPTW 2 & Same as in IPTW 1 + interaction terms for toxicity confounders linearly related to the log-odds. & 0.989 (0.691) & 0.334/5.102 \\ \hline
		IPTW 3 & Same as in IPTW 1 + interaction terms between toxicities and trial linearly related to the log-odds. & 0.977 (0.685) & 0.305/5.245 \\ \hline
		IPTW 4 & Categorical confounders and effect modifier as main effect only; B-spline basis matrix for cubic polynomial splines with three internal knots were used to model the relationship between each continuous MOTox score and the log-odds. & 0.969 (0.813) & 0.306/7.027\\  \hline
		IPTW 5 & Same as in IPTW 1 + interaction terms between toxicities and histological response linearly related to the log-odds & 0.963 (0.694) & 0.279/8.386 \\
		\hline
	\end{tabular*}
\begin{tablenotes}
	\scriptsize
	\item $^1${See Appendix \ref{appD:IPTW} for further details.}
	\end{tablenotes}
\end{table}

\begin{figure}[h!]
	\centering{\includegraphics[width=1\textwidth]{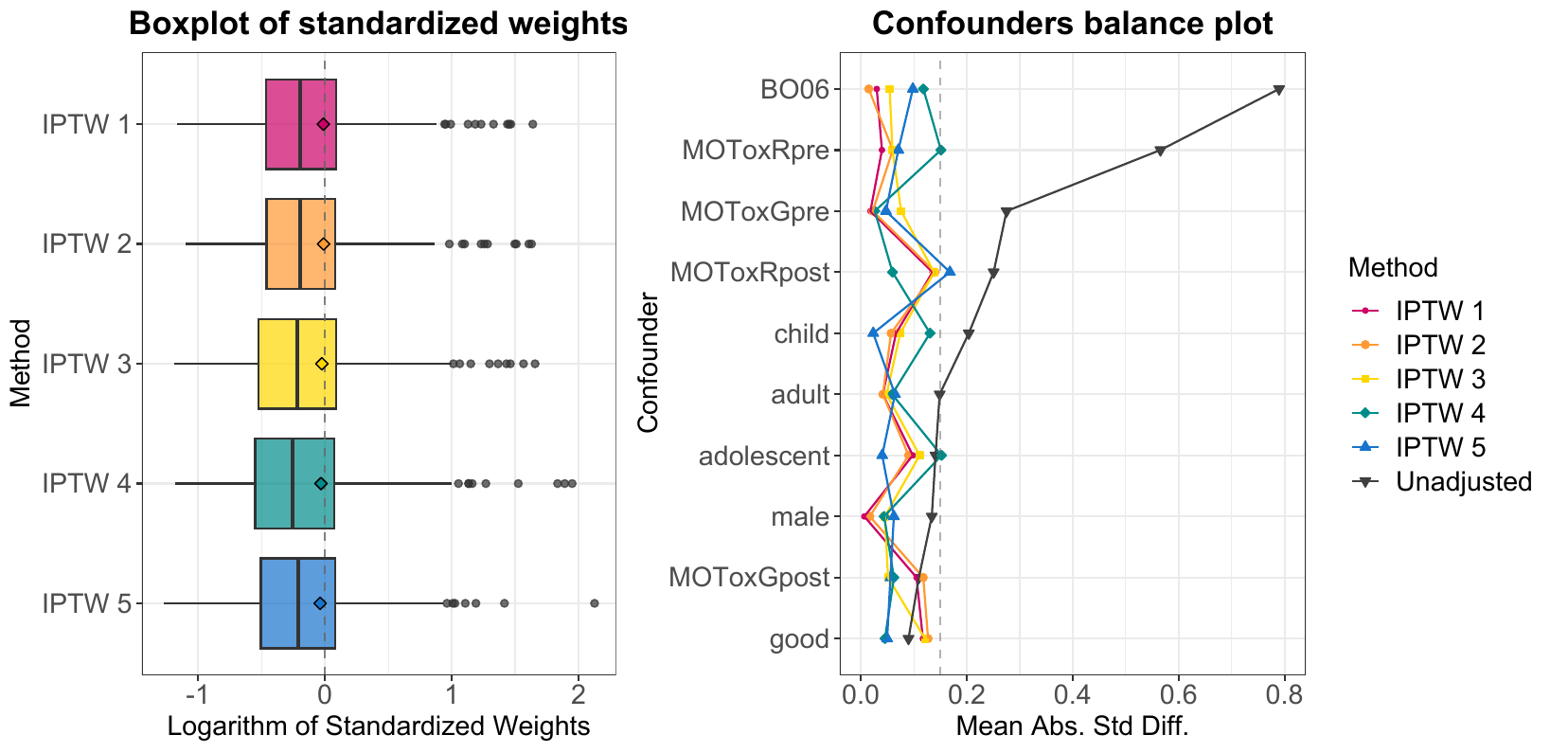}}
	\caption{Diagnostic plots for Inverse Probability of Treatment Weighting (IPTW) performed by using the five different specification methods in Table \ref{tab:IPTW} (purple: \textit{IPTW 1}; orange: \textit{IPTW 2}; yellow:  \textit{IPTW 3}; green:  \textit{IPTW 4}; blue: \textit{IPTW 5}). Left panel: Boxplots of subject-specific stabilized weights $SW_i$ computed via Equations \eqref{eq:sw1} in logarithmic-scale. Diamonds represent the mean values. Right panel: Confounder balance plot. Lines represent the mean absolute standardized differences for each exposure-related confounder according to the four different specification methods (colored lines) and their unadjusted version (black line). \label{fig:diagnostics}}
\end{figure}

The distributions of the stabilized weights (Table \ref{tab:IPTW}; left-panel of Figure \ref{fig:diagnostics}) suggest that there was no evidence of violation of the positivity or misspecification assumptions for IPTW methods 1 and 2 (mean values of about 0.99 without extreme values), whereas methods 3 to 5 presented lower mean values and higher standard deviations.  The same was confirmed by the diagnostics balance plot in the right panel of Figure \ref{fig:diagnostics}. The mean absolute standardized differences for confounders in the unweighted sample (black points) always exceeded those in the weighted samples, and the lowest values were observed for IPTW 1 and 2.  
IPTW model 1 was finally selected due to the lower number of parameters.

\subsection{Estimated causal effects}
The causal parameters $\boldsymbol{\beta}$ in Cox MSM \eqref{eq:MSCM1} were estimated through their consistent parameters $\boldsymbol{\theta}$ in weighted Cox model \eqref{eq:cox1} fitted on the pseudo-population obtained with IPTW 1. Robust standard errors for computing the confidence interval of each coefficient were obtained via the option \texttt{robust=TRUE} in R function \texttt{coxph} \cite{survival}. Estimates were finally compared to the results obtained by fitting a traditional Cox model \cite{cox1972} on the original unweighted population.  

Table \ref{tab:MSMsparam} shows the results in the pseudo-population (parameters $\hat{\boldsymbol{\beta}}$ to the left) and the original population (parameters $\boldsymbol{\hat{\beta}}_{\text{unw}}$ to the right). The difference clearly demonstrates how the latter was affected by the toxicity-treatment-adjustment bias. RDI exposure in the original population was not randomized: the final RDI represented the realization of the treatment trajectory influenced by both the severity of overall toxicity experienced by each patient and physicians' interventions. In the pseudo-population mimicking  the TT, randomization was emulated by adjusting for confounding. Therefore, the results can be interpreted in a causal setting.

\begin{table}[h!]
	\small
	\caption{Estimated parameters $\boldsymbol{\hat{\beta}}$ along with their 95\% Confidence Intervals (CIs) for the Cox MSM in Equation \eqref{eq:MSCM1} and for the corresponding unweighted traditional Cox model.}\label{tab:MSMsparam}
	\centering
	\begin{tabular*}{1\textwidth}{@{\extracolsep\fill}crcrc@{\extracolsep\fill}}
		\hline
		&\multicolumn{2}{@{}c@{}}{\textbf{Cox MSM}} & \multicolumn{2}{@{}c@{}}{\textbf{Unweighted Cox model}} \\ 
		&\multicolumn{2}{@{}c@{}}{\textit{Pseudo-population}} & \multicolumn{2}{@{}c@{}}{\textit{Original population}} \\
		\textbf{Covariate} &  \multicolumn{1}{c}{$\boldsymbol{\hat{\beta}}$} &  \multicolumn{1}{c}{\textit{95\% CIs}}  & \multicolumn{1}{c}{$\boldsymbol{\hat{\beta}}_{\text{unw}}$} &  \multicolumn{1}{c}{\textit{95\% CIs}}    \\
		\hline
		$a=1$ & $-0.536$ & $[-1.031; -0.041]$ & $-0.116$ & $[-0.568; 0.335]$ \\ 
		$a=2$ & $-0.808$ & $[-1.384; -0.231]$ & $-0.359$ & $[-0.844; 0.127]$ \\ 
		$a=1 \times V=1$  & $0.747$ & $[-0.379; 1.873]$ & $-0.006$ & $[-0.997; 0.984]$ \\ 
		$a=2 \times V=1$ & $1.747$ & $[0.554; 2.939]$ & $0.979$ & $[0.035; 1.923]$ \\ 
		$V=1$  & $-1.909$ & $[-2.835; -0.981]$ & $-1.175$ & $[-1.921; -0.429]$ \\ 
		\hline
	\end{tabular*}
\end{table}
\begin{figure}[h!]
	\centering{\includegraphics[width=0.975\textwidth]{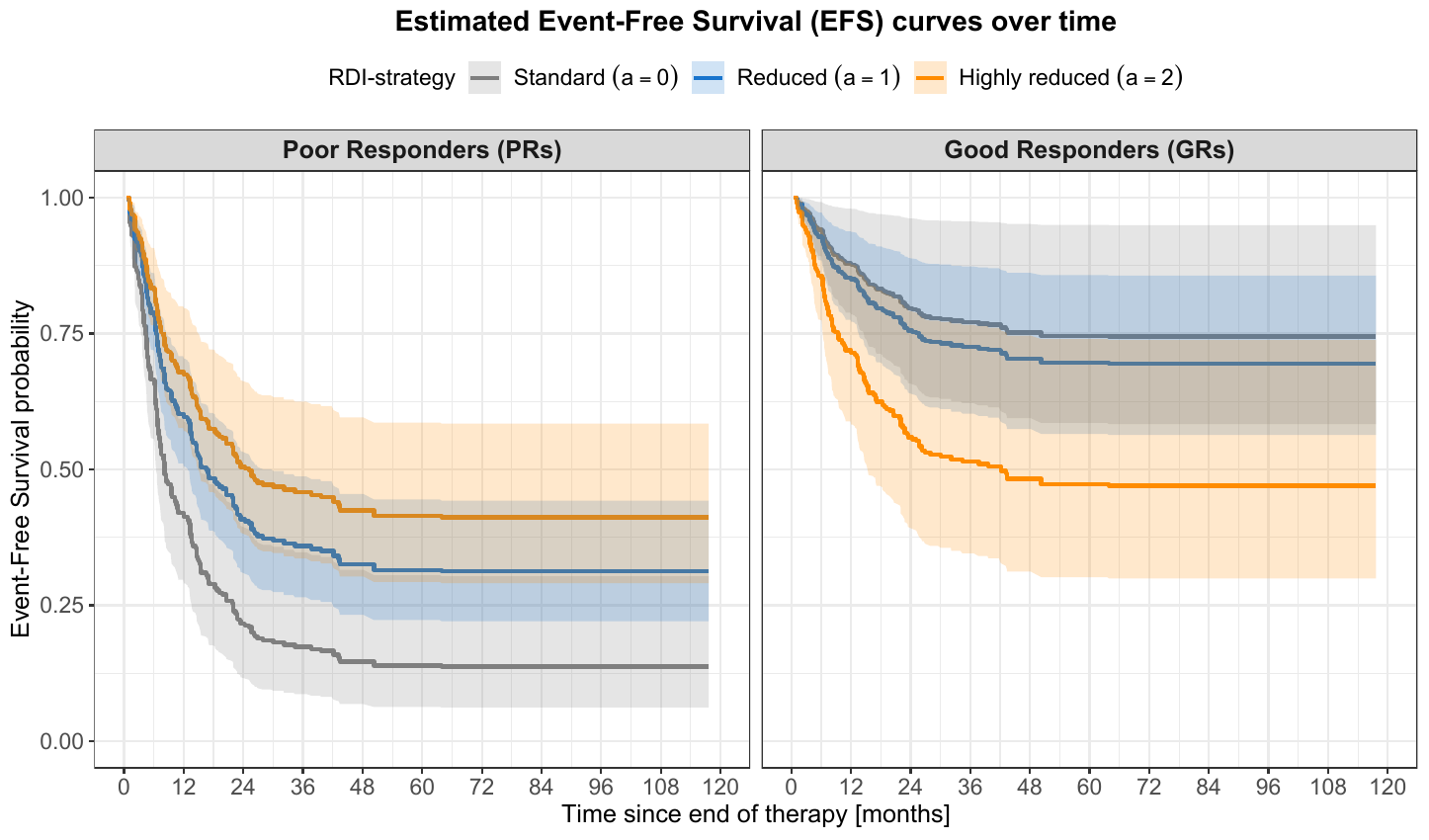}}
	\caption{Estimated Event-Free Survival (EFS) over time $\hat{S}^{a}(t|V=v)$ over time (up to 10 years since end of therapy) for \textit{standard} (gray: $a=0$), \textit{reduced} (blue: $a=1$), and \textit{highly-reduced} RDI (orange: $a=2$) strategies in subgroups of Poor Responders (PRs) (left panel: $v=0$) and Good Responders (GRs) (right panel: $v=1$).}\label{fig:EFScurves}
\end{figure}

Figure \ref{fig:EFScurves} displays the estimated EFS curves $\hat{S}^{a}(t|V=v)$  over time (up to 10 years since end of therapy) for \textit{standard} (gray), \textit{reduced} (blue), and \textit{highly-reduced} (orange) RDI-strategy across \textit{poor} (left panel)  or \textit{good} (right panel) responders.
Results indicate evidence of effect modification: exposures characterized by lower RDI resulted in better EFS in PRs, while, conversely, lower RDI-exposure led to poorer EFS in GRs.

\begin{figure}[h!]
	\centering{\includegraphics[width=0.975\textwidth]{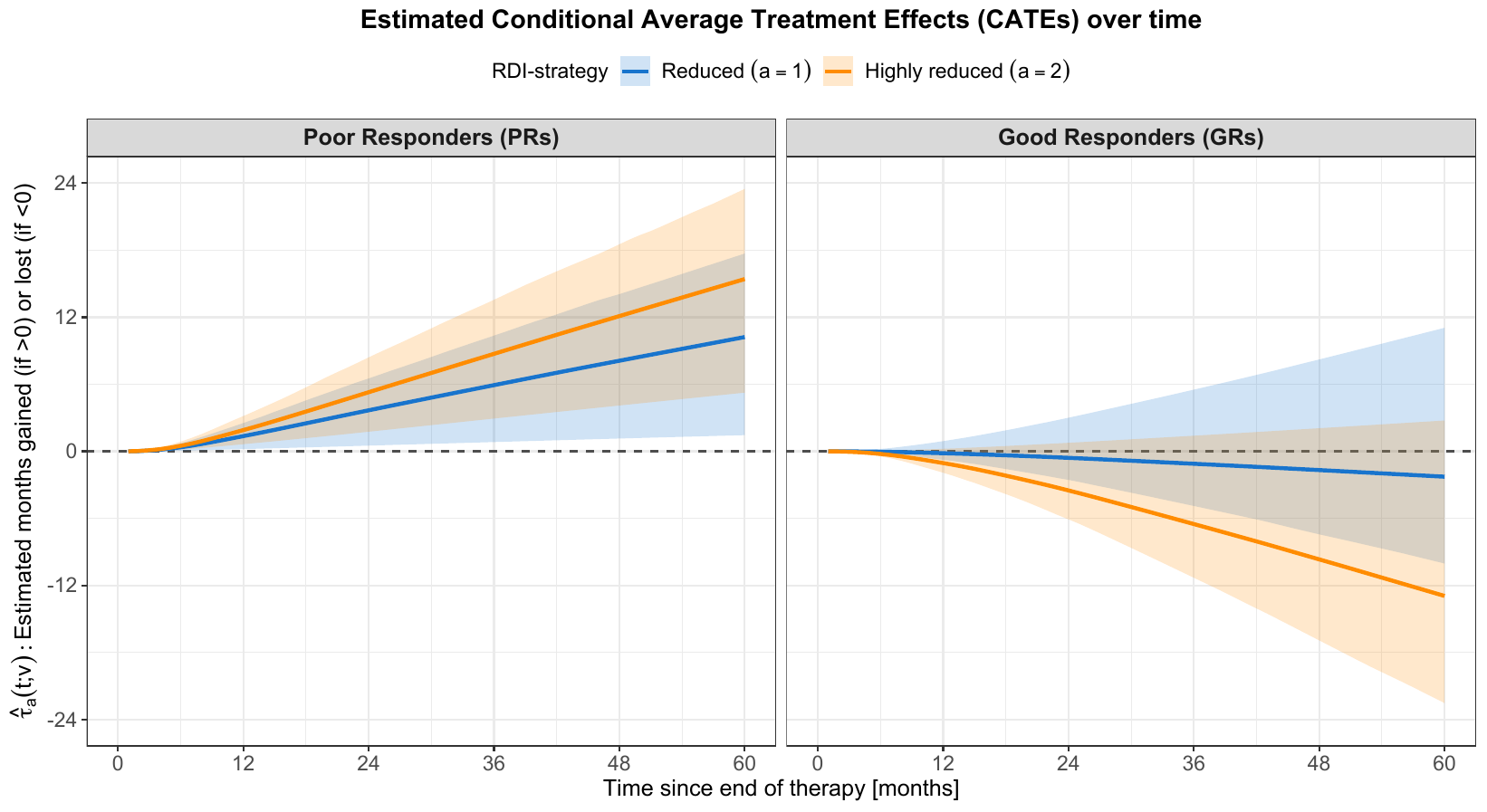}}
	\caption{Estimated Conditional Average Treatment Effects (CATEs) $\hat\tau_{a}(t;v)$ over time (up to 5 years since end of therapy) along with 95\% bootstrap-percentiles CIs for \textit{reduced} (blue: $a=1$) and \textit{highly-reduced} RDI (orange: $a=2$) strategies compared to \textit{standard} in subgroups of Poor Responders (PRs) (left panel: $v=0$) and Good Responders (GRs) (right panel: $v=1$).}\label{fig:cateCIs}
\end{figure}

Figure \ref{fig:cateCIs} shows the estimated CATEs over time (up to 5 years since end of therapy) for \textit{reduced} (blue) and \textit{highly-reduced} (orange) RDI-strategy (compared to \textit{standard}) across patients with \textit{poor} or  \textit{good} HRe, along with the estimated 95\% bootstrap CIs.
The CATE trends differ between PR and GR subgroups due to the heterogeneous effect of RDI reductions. GRs (right panel) exhibited a trend towards a clinical disadvantage resulting from reduced RDI, especially for \textit{high-reduced} strategy. Conversely, PRs (left panel) showed a clinically relevant benefit from reducing RDI, meaning that the intrinsic nature of PRs induced resistance to chemotherapy. In particular, 5-year estimated CATEs ($t=60$ months) were $\hat\tau_{1}(60;0)=10.2$ (95\% $CI = [1.5,17.7]$) and $\hat\tau_{2}(60;0)=15.4$ (95\% $CI = [5.2,23.5]$), indicating an average gain of 10.2 and 15.4 months for \textit{reduced} and \textit{highly-reduced} exposure, respectively. 

A possible clinical explanation for these results might be due to the effect of chemotherapy on non-cancerous cells. By targeting a broad spectrum of cells, chemotherapy also damages the processes and mechanisms of the immune system that can detect and kill cancer cells. While in GRs this negative effect may be largely offset by the efficacy of the tumour therapy, in PRs chemotherapy is less effective due to the impact on the immune system so a higher RDI may be detrimental to survival.

\vspace{5mm}

\section{Discussion}\label{s:disc}

Motivated by a sharp yet delicate clinical question, this paper introduces a novel approach to mimic a hypothetical target trial using RCT data with interventions. The final aim was to investigate the effect of reductions in RDI on EFS in patients with osteosarcoma, with a focus on subgroups of poor  and good responders. Chemotherapy administration data in osteosarcoma from BO03 and BO06 RCTs were analysed. IPTW was first used to transform the original selected population into a pseudo-population emulating the randomized cohort of the TT. Then, Cox MSM with effect modification was employed to compared the effects of RDI reductions ranging from 15\% to 30\% (\textit{reduced} exposure) or above 30\% (\textit{highly-reduced} exposure) to the \textit{standard} RDI of EOI tratment (structured in 6 cycles of 3-weekly CDDP+DOX) in both PRs and GRs. CATEs were finally measured as the contrast between the RMST of \textit{reduced/highly-reduced} RDI-strategy and that of the \textit{standard} one. The 95\% CIs for CATEs were obtained using a novel IPTW-based bootstrap procedure while preserving the sizes of sub-cohorts.

Considering the data complexity and the underlying causal assumptions, a note of caution is required, as it needs to encompass all aspects of the chemotherapy process.
First, exposure and outcome must be properly defined to guarantee the consistency assumption. Then, pre- and post-assigned confounders must be carefully identified to satisfy the assumption of no unmeasured confounding. During this process, it is imperative to notice that assignment of dose reductions or delays in chemotherapy administration was determined not by individual toxicities but by the overall toxic burden for each patient. Therefore, pre- and post-operative side-effects data were summarized using the new Multiple Overall Toxicity (MOTox) approach \cite{spreaficoBMJOpen} for both \textit{rule-specific} and \textit{generic} toxicity. This novel analytical strategy allowed (i) to reduce the number of possible confounders combinations dealing with non-positivity and highly-correlated data, and (ii) to meet the clinical rationale of tailoring treatment according to the patient's overall toxic burden in the presence of multiple toxic side effects. Third, different weighting models have to be compared in order to preserve positivity and guarantee a correct IPTW specification. Finally, an outcome model to address the research question at hand must be correctly specified. This led to the definition of a Cox MSM with effect modification given by HRe, which represents the causal RDI analogue of the ITT landmark Cox model presented in Lewis \textit{et al.} (2007)\cite{lewis2007}.

The first significant contribution of this study is its innovative use of TT emulation to address the research question. The results revealed evidence for effect modifications by HRe, as increasing RDI-reductions caused two opposite trends for PRs and GRs. Specifically, higher RDI reductions led to improved EFS in PRs but worsened EFS in GRs. Estimated CATEs highlighted that PRs can significantly benefit from reduced RDI, due to their intrinsic resistance to chemotherapy. Future studies should investigate this phenomenon. 
Evidence of effect modification can be exploited for establishing new treatment guidelines tailored to specific patient subgroups that could benefit from modified treatment strategies. The findings of this study suggested that guidelines should recommend reduced dose intensity for PRs, while the situation in GRs is less severe and may warrant a case-by-case decision.

Furthermore, the proposed TT emulation approach enabled the identification of potential pitfalls in a naive RDI-based analysis of chemotherapy data. When the ITT model from Lewis \textit{et al.} (2007)\cite{lewis2007} was adapted into the traditional Cox models fitted on the unweighted original population by neglecting the influence of toxicities or other confounding factors, the results were  influenced by the presence of the \textit{toxicity-treatment-adjustment} bias. By employing an IPTW-based Cox MSM, this study eliminated the feedback loop between side effects and treatment adjustments. This resulted in unbiased estimates of the impact of RDI reductions on EFS within the two subgroups and provided a more accurate depiction of the effects of low-intensity regimens.

The third significant contribution of our study is to have demonstrated how existing RCT data can be effectively repurposed for additional retrospective analyses, extending beyond the intended scope of the original studies. Our proposed analytical approach possesses the versatility to be adapted and applied to various cancer-related investigations. These investigations might encompass diverse types of treatments (e.g., immunotherapies or molecularly targeted agents) with its unique set of side effects to study how reductions in treatment intensity influence the outcome of interest, possibly within specific subgroups of interest. This would require a detailed protocol and close collaboration with medical staff to identify patient’s clinical history, relevant side effects, and treatment-related factors.

Finally, a novel generalized bootstrap procedure was introduced to compute confidence intervals for CATEs. This approach diverged from typical random sampling in two key ways: (i) it sampled from each sub-cohort characterized by various combinations of strategies and effect modifier levels while maintaining the sub-sample sizes, and (ii) it employed unequal IPTW-based probability sampling \cite{DONALD2014,genbootstrap}.
By employing this procedure, all sub-cohorts were adequately represented in the generalized bootstrap samples, avoiding estimation issues due to missing observations. Moreover, subjects with oversized weights were sampled more frequently, allowing for a more thorough exploration of the uncertainty associated with them \cite{genbootstrap}.
This approach heavily relies on the estimated stabilized weights. Therefore, it is crucial that the theoretical assumption of no unmeasured confounding holds, and the IPTW model must be correctly specified.

\section{Conclusions}\label{s:conclusions}
This work has introduced an innovative and comprehensive analysis of chemotherapy administration RCT data, aimed at addressing specific clinical questions related to the reduction of RDI within subgroups of patients with osteosarcoma. The study is complemented by tutorial-like explanations which provide insights into the inherent challenges in this scenario and the novel problem-solving strategies proposed. Furthermore, this study has emphasized the critical role of toxicities in this context and illustrated the detrimental consequences of neglecting them in the analyses.

To the best of our knowledge, no other studies have employed the principle of TT emulation to address this important research question in the oncological field. The developed approach offers several advantages, including (i) accounting for all the unique aspects of chemotherapy, (ii) mitigating the \textit{toxicity-treatment-adjustment} bias, and (iii) effectively repurposing existing RCT data for additional retrospective analyses extending beyond the intended scope of the original trials.

\subsubsection*{Acknowledgments}
The authors would like to express their gratitude to Dr. Carlo Lancia and Dr. Cristian Spitoni for their valuable preliminary analysis, which provided the foundation for this study. Special thanks to the Medical Research Council in London for generously sharing the dataset used in this research, and to Dr. Jakob Anninga, MD for providing clinical insights. F.I. has been supported by MUR, grant Dipartimento di Eccellenza 2023-2027.

\subsubsection*{Data \& code availability} Original data are not publicly available due to privacy restrictions. Access to the full datasets can be requested to MRC Clinical Trials Unit at UCL, Institute of Clinical Trials and Methodology, UCL, London. 
Software in the form of \texttt{R} code,\cite{R} together with a toy sample input dataset and complete
documentation is available here: \url{https://github.com/mspreafico/TTEcausalRDI}

\newpage
\renewcommand{\thesection}{\Alph{section}}

\Large \section*{Appendix} \normalsize

\renewcommand\thesubsection{A}
\subsection{CTCAE grades specification for \textit{rule-specific} and \textit{generic} toxicities}\label{appA:ctcae}

In both BO03 and BO06 studies, toxic side effects were recorded using the Common Terminology Criteria for Adverse Events Version 3 (CTCAE v3.0) \cite{ctcae3}, with grades ranging from 0 (none) to 4 (life-threatening). Table \ref{tab:ctcae} reports the CTCAE-grades for \textit{rule-specific} toxicities (i.e., \textit{leucopenia}, \textit{thrombocytopenia}, \textit{oral mucositis}, \textit{ototoxicity}, \textit{cardiotoxicity} and \textit{neurotoxicity}) and \textit{generic} ones (i.e., \textit{nausea/vomiting} and \textit{infections}).

\begin{sidewaystable}
	\renewcommand\thetable{A1}
	\caption{Toxicity coding based on Common Terminology Criteria for Adverse Events (CTCAE) v3.0\cite{ctcae3} for \textit{rule-specific} (i.e., leucopenia, thrombocytopenia, oral mucositis, ototoxicity, cardiotoxicity and neurotoxicity) and \textit{generic} (i.e., nausea/vomiting and infections) toxicities.\label{tab:ctcae}}
	\begin{tabular*}{\textheight}{@{\extracolsep\fill}p{2.5cm}p{2.2cm}p{2.8cm}p{2.8cm}p{3.1cm}p{2.8cm}}%
		\hline
		\textbf{Toxicity} & \textbf{Grade 0} & \textbf{Grade 1} &\textbf{Grade 2} &\textbf{Grade 3} &\textbf{Grade 4} \\
		\hline \hline
		\textit{\textbf{Rule-specific}} &&&&&\\ 
		\textit{Leucopenia } & $\ge 4.0 \times 10^9/L$ & $[3.0-4.0) \times 10^9/L$ & $[2.0-3.0) \times 10^9/L$ & $[1.0-2.0) \times 10^9/L$ & $< 1.0 \times 10^9/L$  \\ \hline
		\textit{Thrombocytonpenia} & $\ge 100 \times 10^9/L$  & $[75-100) \times 10^9/L$ & $[50-75) \times 10^9/L$ & $[25-50) \times 10^9/L$ & $<25\times 10^9/L$ \\ \hline
		\textit{Oral Mucositis} & No change & Soreness or erythema & Ulcers: can eat solid & Ulcers: liquid diet only & Alimentation not possible\\ \hline
		\textit{Cardiac toxicity} & No change & Sinus tachycardia & Unifocal PVC\footnotemark[1] arrhythmia & Multifocal PVC\footnotemark[1] & Ventricular tachycardia\\ \hline
		\textit{Ototoxicity} & No change & Slight hearing loss & Moderate hearing loss & Major hearing loss & Complete hearing loss\\ \hline
		\textit{Neurological \newline toxicity} & None & Paraesthesia & Severe paraesthesia & Intolerable paraesthesia & Paralysis\\
		\hline \hline
		\textit{\textbf{Generic}} &&&&&\\ 
		\textit{Nausea/Vomiting} & None & Nausea & Transient vomiting & Continuative vomiting  & Intractable vomiting\\  \hline
		\textit{Infection} & None & Minor infection & Moderate infection  & Major infection & Major infection with hypotension\\
		\hline
	\end{tabular*}
\scriptsize
	$^1${PVC = Premature Ventricular Contraction}
\end{sidewaystable}

\renewcommand\thesubsection{B}
\subsection{Assumptions for causal inference through marginal structural models}\label{appB:CIassumptions} 
The four main assumptions for causal inference with Marginal Structural Models (MSMs) through Inverse Probability of Treatment Weighting (IPTW) are here discussed.\\

\textbf{\textit{Exchangeability}} (or conditional exchangeability) implies the well-known assumption of \textit{no unmeasured confounding} \cite{colehernan2008}. It states that exposure allocation is independent of the potential outcomes conditional on confounders \cite{colehernan2008}, that is: $$T^{a} \independent A | \boldsymbol{L}.$$ 

In the absence of randomization, as is the case in observational studies or RCTs with interventions, it is not possible to test for exchangeability. In such situations, expert knowledge becomes essential for identifying an adequate set of joint predictors of exposure and outcome. These predictors should be chosen in a way that, within their respective levels, any associations between exposure and outcome resulting from shared underlying causes are effectively controlled for \cite{colehernan2008}.\\

\textbf{\textit{Consistency}} means that the outcome observed for each individual is  the counterfactual outcome under the observed treatment history, that is: 
$$ T^{a} = T_i \quad \text{ for every individual $i$ with } A_i= a.$$
This assumption is violated in the presence of misclassification bias \cite{ravani2017} and has two requirements \cite{cibook2020}:
\begin{enumerate}
	\item [i.]	the exposure must be properly defined so that the counterfactual outcomes are well-defined (this implies that a specific exposure may be hypothetically assigned to a subject  exposed to a different level);
	\item [ii.] a link between counterfactuals and observed data is reasonable in the context under study (this means that the equality should be valid for at least some individuals).
\end{enumerate}
Although consistency can not be empirically verified, it is assumed to be plausible in (observational) studies of medical treatments, since it may be possible to change an individual's treatment status \cite{cole2009epi}.\\

\textbf{\textit{Positivity}} states that there is a non-zero (i.e., positive) probability of receiving every level of exposure for every
combination of covariates that occur among individuals in the population \cite{colehernan2008}. In the context under analysis, this corresponds to
$$ \Pr\left(A_i = a \big| \boldsymbol{L}_i=\boldsymbol{l}, V_i=v \right) > 0 \quad \forall a,\boldsymbol{l},v.$$
If this assumption is violated, then the weights in IPTW in Equation \eqref{eq:sw1} are undefined leading to biased estimates of the causal effect. 

If a subject cannot be exposed to one or more levels of the confounders (e.g., it cannot be treated in the presence of recommendations from guidelines or established contraindications), then positivity is violated due to a \textit{structural} zero probability of receiving the specific exposure. A solution is to restrict the inference to the subset with a positive probability of exposure, whenever possible \cite{cole2009epi}. 	Even in the absence of structural zeros, \textit{random} zeros may occur by chance due to small sample sizes  or highly stratified data by numerous confounders.
The inclusion of weak or highly-stratified confounders can provide a better confounding adjustment but may cause severe non-positivity, increasing the bias and variance of the estimated effect. 
An indication of non-positivity may be the presence of estimated weights with the mean far from one or very extreme values \cite{cole2009epi}.\\

\textbf{\textit{No misspecification of both weighting and outcome models}}
means that both the weighting model for IPTW and the structural outcome model, which links the outcome to the exposure history, must be correctly specified. This assumption has similar roots in all statistical models \cite{ravani2017}, as model misspecification leads to instability in the Cox MSM estimates \cite{karim2014,karim2017}.

Since the presence of estimated stabilized weights with the mean far from one or with extreme values suggest possible violation of positivity or misspecification of the weight model \cite{cole2009epi}, proper model specifications can be checked by exploring the distribution of weights \cite{colehernan2008}.
In addition, quantitative (e.g., weighted standardized difference to compare means or prevalences) and qualitative graphical methods can be used to assess whether measured covariates are balanced between treatment groups in the weighted sample \cite{austinstuart2015}.

\renewcommand\thesubsection{C}
\subsection{Flowchart BO03/BO06 cohort selection}\label{appC:cohort}
\begin{figure}[h!]
	\renewcommand\thefigure{C1}
	\centering{\includegraphics[width=0.9\textwidth]{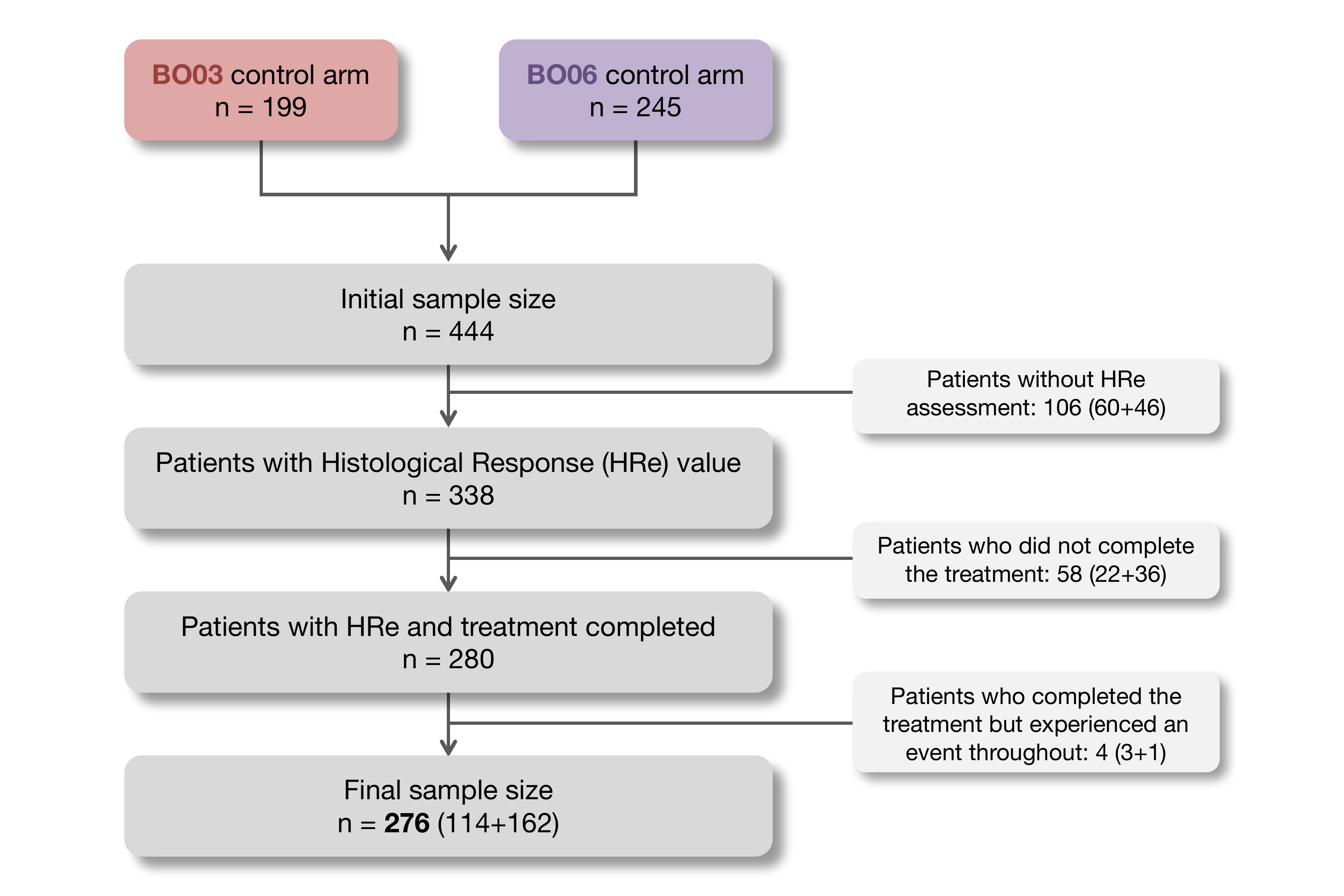}}
	\caption{Flowchart of cohort selection.\label{fig:consort}}
\end{figure}

Figure \ref{fig:consort} displays the consort diagram related to the final cohort of 276 patients (114 and 162 from BO03 and BO06, respectively) included in the analyses.

\renewcommand\thesubsection{D}
\subsection{Examining denominator models for IPTW}\label{appD:IPTW}

Different specifications of the subject-specific stabilized weights are investigated in order to check whether and which models best satisfied \textit{positivity} and \textit{no misspecification}.
Multinomial logistic regression models are used for both numerators and denominators of $SW_i$ in Equation \eqref{eq:sw1}. 

The numerators $\Pr\left(A_i | V_i \right)$, i.e., the probability that a subject $i$ received exposure $A_i$ given his/her histological response $V_i$, are modelled as: 
\begin{align*}
	\Pr\left(A_i = a \big| V_i\right) &= \frac{\exp\left(\alpha_{0a}  + \alpha_{1a} \cdot \texttt{GR}_i\right)}{1+ \sum_{j=1}^2 \exp\left(\alpha_{0j}  + \alpha_{1j} \cdot \texttt{GR}_i\right)}  \qquad a = 1,2;\\
	\Pr\left(A_i = 0 \big| V_i\right) &= \frac{1}{1+ \sum_{j=1}^2 \exp\left(\alpha_{0j}  + \alpha_{1j} \cdot \texttt{GR}_i\right)},
\end{align*}
where variable \texttt{GR} is the dummy variable for good responders created from categorical variable \texttt{HRe} (\textit{poor}; \textit{good}).

The denominator $P\left(A_i \big|  \boldsymbol{L}_i, V_i\right)$ is the probability that the subject received exposure $A_i$ given relative confounders $\boldsymbol{L}_i$ and effect modifier $V_i$. Five different options are modelled as: 
\begin{align*}
	\Pr\left(A_i = a \big| \boldsymbol{L}_i, V_i\right) &= \frac{\exp\left(\eta_{ia}\right)}{1+ \sum_{j=1}^2 \exp\left(\eta_{ij}\right)}  \qquad a = 1,2;\\
	\Pr\left(A_i = 0 \big| \boldsymbol{L}_i, V_i \right) &= \frac{1}{1+ \sum_{j=1}^2  \exp\left(\eta_{ij}\right)}.
\end{align*}
by considering five different linear predictors $\eta_{ia}$ in terms of confounding features.
\begin{itemize}
	\item \textbf{IPTW 1}: categorical/binary confounding covariates and binary effect modifier are included as a main effect only and the MOTox scores are assumed linearly related to the log-odds:
	\begin{align*}
		\eta_{ia} = \gamma_{0a} &+ \gamma_{1a}\cdot\texttt{BO06}_i +
		\gamma_{2a} \cdot \texttt{adolescent}_i +
		\gamma_{3a} \cdot \texttt{adult}_i + \gamma_{4a} \cdot \texttt{male}_i  + \\ 
		& +\gamma_{5a}  \cdot \texttt{MOTox}_{i,gen,pre} + \gamma_{6a}  \cdot \texttt{MOTox}_{i,rule,pre} +\\ 
		& +
		\gamma_{7a}  \cdot \texttt{MOTox}_{i,gen,post} + \gamma_{8a}  \cdot \texttt{MOTox}_{i,rule,post} + \gamma_{9a} \cdot \texttt{GR}_i.
	\end{align*}
	\item \textbf{IPTW 2}: same as in IPTW 1 + interaction terms for toxicity confounders linearly related to the log-odds:
	\begin{align*}
		\eta_{ia} = \gamma_{0a} &+ \gamma_{1a}\cdot\texttt{BO06}_i +
		\gamma_{2a} \cdot \texttt{adolescent}_i +
		\gamma_{3a} \cdot \texttt{adult}_i + \gamma_{4a} \cdot \texttt{male}_i  + \\ 
		& +\gamma_{5a}  \cdot \texttt{MOTox}_{i,gen,pre} + \gamma_{6a}  \cdot \texttt{MOTox}_{i,rule,pre} +\\ 
		& +
		\gamma_{7a}  \cdot \texttt{MOTox}_{i,gen,post} + \gamma_{8a}  \cdot \texttt{MOTox}_{i,rule,post} + \gamma_{9a} \cdot \texttt{GR}_i +\\
		& + \gamma_{10a}  \cdot \texttt{MOTox}_{i,gen,pre} \cdot \texttt{MOTox}_{i,rule,pre} + \gamma_{11a}  \cdot \texttt{MOTox}_{i,gen,post} \cdot \texttt{MOTox}_{i,rule,post}.
	\end{align*}
	\item \textbf{IPTW 3}: same as in IPTW 1 + interaction terms between toxicities and trial linearly related to the log-odds:
	\begin{align*}
		\eta_{ia} = \gamma_{0a} &+ \gamma_{1a}\cdot\texttt{BO06}_i +
		\gamma_{2a} \cdot \texttt{adolescent}_i +
		\gamma_{3a} \cdot \texttt{adult}_i + \gamma_{4a} \cdot \texttt{male}_i  + \\ 
		& +\gamma_{5a}  \cdot \texttt{MOTox}_{i,gen,pre} + \gamma_{6a}  \cdot \texttt{MOTox}_{i,rule,pre} +\\ 
		& +
		\gamma_{7a}  \cdot \texttt{MOTox}_{i,gen,post} + \gamma_{8a}  \cdot \texttt{MOTox}_{i,rule,post} + \gamma_{9a} \cdot \texttt{GR}_i +\\
		& + \gamma_{10a}  \cdot\texttt{trial}_i \cdot \texttt{MOTox}_{i,gen,pre}  + \gamma_{11a} \cdot\texttt{trial}_i \cdot \texttt{MOTox}_{i,rule,pre}  + \\ &+ \gamma_{12a} \cdot\texttt{trial}_i  \cdot \texttt{MOTox}_{i,gen,post} + \gamma_{13a} \cdot\texttt{trial}_i \cdot \texttt{MOTox}_{i,rule,post}.
	\end{align*}
	This choice is motivated by the statistically different distributions of the MOTox scores in BO03 and BO06 trials (see Table \ref{tab:descriptive}).\\
	\item \textbf{IPTW 4}: categorical/binary confounding covariates and binary effect modifier are included as a main effect only;
	B-spline basis matrix for cubic polynomial splines with three internal knots are used to model the relationship between each continuous MOTox score and the log-odds: 
	\begin{align*}
		\eta_{ia} = \gamma_{0a} &+ \gamma_{1a}\cdot\texttt{BO06}_i +
		\gamma_{2a} \cdot \texttt{adolescent}_i +
		\gamma_{3a} \cdot \texttt{adult}_i + \gamma_{4a} \cdot \texttt{male}_i  + \\ 
		&+ 
		\boldsymbol\gamma_{5a}^T \boldsymbol{B}\left(\texttt{MOTox}_{i,gen,pre}\right) + \boldsymbol\gamma_{6a}^T \boldsymbol{B}\left(\texttt{MOTox}_{i,rule,pre}\right) +\\
		&+    \boldsymbol\gamma_{7a}^T\boldsymbol{B}\left(\texttt{MOTox}_{i,gen,post}\right) + \boldsymbol\gamma_{8a}^T \boldsymbol{B}\left(\texttt{MOTox}_{i,rule,post}\right)  + \gamma_{9a} \cdot \texttt{GR}_i.
	\end{align*}
	\item \textbf{IPTW 5}: same as in IPTW 1 + interaction terms between toxicities and HRe linearly related to the log-odds:
	\begin{align*}
		\eta_{ia} = \gamma_{0a} &+ \gamma_{1a}\cdot\texttt{BO06}_i +
		\gamma_{2a} \cdot \texttt{adolescent}_i +
		\gamma_{3a} \cdot \texttt{adult}_i + \gamma_{4a} \cdot \texttt{male}_i  + \\ 
		& +\gamma_{5a}  \cdot \texttt{MOTox}_{i,gen,pre} + \gamma_{6a}  \cdot \texttt{MOTox}_{i,rule,pre} +\\ 
		& +
		\gamma_{7a}  \cdot \texttt{MOTox}_{i,gen,post} + \gamma_{8a}  \cdot \texttt{MOTox}_{i,rule,post} + \gamma_{9a} \cdot \texttt{GR}_i +\\
		& + \gamma_{10a}  \cdot\texttt{GR}_i \cdot \texttt{MOTox}_{i,gen,pre}  + \gamma_{11a} \cdot\texttt{GR}_i \cdot \texttt{MOTox}_{i,rule,pre}  + \\ &+ \gamma_{12a} \cdot\texttt{GR}_i  \cdot \texttt{MOTox}_{i,gen,post} + \gamma_{13a} \cdot\texttt{GR}_i \cdot \texttt{MOTox}_{i,rule,post}.
	\end{align*}
	
	Variable \texttt{BO06} is the dummy variable created from categorical variable \texttt{trial} (\textit{BO03}; \textit{BO06}). Variables \texttt{adolescent} and \texttt{adult} are the dummy variables created from categorical variable \texttt{age} (\textit{child}; \textit{adolescent}; \textit{adult}). Variable \texttt{male} is the dummy variable created from categorical variable \texttt{gender} (\textit{female}; \textit{male}).
	
\end{itemize}

\renewcommand\thesubsection{A}

\begin{figure}[h!]
	\renewcommand\thefigure{A.1}
\end{figure}


\small
\begin{thebibliography}{10}
	\providecommand{\doi}[1]{\url{https://doi.org/#1}}
	
	\bibitem{smeland2019}
	Smeland S, Bielack SS, Whelan J, Bernstein M, Hogendoorn P, Krailo MD, et~al.
	\newblock {Survival and prognosis with osteosarcoma: outcomes in more than 2000
		patients in the EURAMOS-1 (European and American Osteosarcoma Study) cohort}.
	\newblock European Journal of Cancer. 2019;109:36--50.
	\newblock \doi{10.1016/j.ejca.2018.11.027}.
	
	\bibitem{ritter2010}
	Ritter J, Bielack SS.
	\newblock {Osteosarcoma}.
	\newblock Annals of Oncology. 2010 10;21(suppl 7):vii320--vii325.
	\newblock \doi{10.1093/annonc/mdq276}.
	
	\bibitem{anninga2011}
	Anninga JK, Gelderblom H, Fiocco M, Kroep JR, Taminiau AHM, Hogendoorn PCW,
	et~al.
	\newblock {Chemotherapeutic adjuvant treatment for osteosarcoma: Where do we
		stand?}
	\newblock European Journal of Cancer. 2011;47(16):2431--2445.
	\newblock \doi{10.1016/j.ejca.2011.05.030}.
	
	\bibitem{bishop2016}
	Bishop MW, Chang YC, Krailo MD, Meyers PA, Provisor AJ, Schwartz CL, et~al.
	\newblock {Assessing the Prognostic Significance of Histologic Response in
		Osteosarcoma: A Comparison of Outcomes on CCG-782 and INT0133-A Report From
		the Children's Oncology Group Bone Tumor Committee}.
	\newblock Pediatric blood \& cancer. 2016;63(10):1737--1743.
	\newblock \doi{10.1002/pbc.26034}.
	
	\bibitem{lancia2019method}
	Lancia C, Anninga J, Sydes MR, Spitoni C, Whelan J, Hogendoorn PCW, et~al.
	\newblock {Method to measure the mismatch between target and achieved received
		dose intensity of chemotherapy in cancer trials: a retrospective analysis of
		the MRC BO06 trial in osteosarcoma}.
	\newblock BMJ open. {2019};9(5).
	\newblock \doi{10.1136/bmjopen-2018-022980}.
	
	\bibitem{lewis2007}
	Lewis IJ, Nooij MA, Whelan J, Sydes MR, Grimer R, Hogendoorn PCW, et~al.
	\newblock {Improvement in Histologic Response But Not Survival in Osteosarcoma
		Patients Treated With Intensified Chemotherapy: A Randomized Phase III Trial
		of the European Osteosarcoma Intergroup}.
	\newblock JNCI: Journal of the National Cancer Institute. 2007;99(2):112--128.
	\newblock \doi{10.1093/jnci/djk015}.
	
	\bibitem{gupta2011}
	Gupta SK.
	\newblock {Intention-to-treat concept: A review}.
	\newblock Perspectives in clinical research. 2011;2(3):109--112.
	\newblock \doi{10.4103/2229-3485.83221}.
	
	\bibitem{Smith2021}
	Smith VA, Coffman CJ, Hudgens MG.
	\newblock Interpreting the results of intention-to-treat, per-protocol, and
	as-treated analyses of clinical trials.
	\newblock Journal of the American Medical Association. 2021;326(5):433--434.
	\newblock \doi{10.1001/jama.2021.2825}.
	
	\bibitem{souhami1997}
	Souhami RL, Craft AW, {Van der Eijken} JW, Nooij M, Spooner D, Bramwell VH,
	et~al.
	\newblock {Randomised trial of two regimens of chemotherapy in operable
		osteosarcoma: a study of the European Osteosarcoma Intergroup}.
	\newblock The Lancet. 1997;350(9082):911--917.
	\newblock \doi{10.1016/S0140-6736(97)02307-6}.
	
	\bibitem{lancia2019}
	Lancia C, Spitoni C, Anninga J, Whelan J, Sydes MR, Jovic G, et~al.
	\newblock {Marginal structural models with dose-delay joint-exposure for
		assessing variations to chemotherapy intensity}.
	\newblock Statistical Methods in Medical Research. 2019;28(9):2787--2801.
	\newblock PMID: 29916309. \doi{10.1177/0962280218780619}.
	
	\bibitem{lancia2019novel}
	Lancia C, Anninga J, Sydes MR, Spitoni C, Whelan J, Hogendoorn PCW, et~al.
	\newblock {A novel method to address the association between received dose
		intensity and survival outcome: benefits of approaching treatment
		intensification at a more individualised level in a trial of the European
		Osteosarcoma Intergroup}.
	\newblock Cancer chemotherapy and pharmacology. {2019};83(5):951--962.
	\newblock \doi{10.1007/s00280-019-03797-3}.
	
	\bibitem{RDIpaper}
	Hryniuk WM, Goodyear M.
	\newblock {The calculation of received dose intensity}.
	\newblock Journal of Clinical Oncology. 1990;8(12):1935--1937.
	\newblock \doi{10.1200/JCO.1990.8.12.1935}.
	
	\bibitem{spreaficoSMAP}
	Spreafico M, Ieva F, Fiocco M.
	\newblock Modelling time-varying covariates effect on survival via functional
	data analysis: application to the {MRC} {BO06} trial in osteosarcoma.
	\newblock {Statistical Methods \& Applications}. 2023;32:271--298.
	\newblock \doi{10.1007/s10260-022-00647-0}.
	
	\bibitem{cox1972}
	Cox DR.
	\newblock {Regression models and life-tables}.
	\newblock Journal of the Royal Statistical Society. 1972;34(2):187--220.
	
	\bibitem{Therneau2010-cf}
	Therneau T, Grambsch P.
	\newblock {Modeling survival data: Extending the Cox model}.
	\newblock 1st ed. Statistics for Biology and Health. New York, NY: Springer;
	2010.
	
	\bibitem{Kleinbaum2016-xr}
	Kleinbaum DG, Klein M.
	\newblock {Survival Analysis}.
	\newblock Statistics for Biology and Health. New York, NY: Springer; 2016.
	
	\bibitem{hernan2016}
	Hern{\'a}n MA, Robins JM.
	\newblock {Using big data to emulate a target trial when a randomized trial is
		not available}.
	\newblock American Journal of Epidemiology. 2016;183(8):758--764.
	\newblock \doi{10.1093/aje/kwv254}.
	
	\bibitem{lewis2000}
	Lewis IJ, Weeden S, Machin D, Stark D, Craft AWa.
	\newblock {Received Dose and Dose-Intensity of Chemotherapy and Outcome in
		Nonmetastatic Extremity Osteosarcoma}.
	\newblock Journal of Clinical Oncology. 2000;18(24):4028--4037.
	\newblock \doi{10.1200/JCO.2000.18.24.4028}.
	
	\bibitem{cibook2020}
	Hern\'{a}n M, Robins J.
	\newblock {Causal Inference: What If}.
	\newblock Chapman \& Hall/CRC BRC, editor; 2020.
	
	\bibitem{efron1979}
	Efron B.
	\newblock {Bootstrap Methods: Another Look at the Jackknife}.
	\newblock The Annals of Statistics. 1979;7(1):1--26.
	
	\bibitem{efrontib1994}
	Efron B, Tibshirani RJ.
	\newblock An introduction to the bootstrap.
	\newblock Chapman and Hall/CRC; 1994.
	
	\bibitem{DONALD2014}
	Donald SG, Hsu YC.
	\newblock Estimation and inference for distribution functions and quantile
	functions in treatment effect models.
	\newblock Journal of Econometrics. 2014;178:383--397.
	\newblock \doi{https://doi.org/10.1016/j.jeconom.2013.03.010}.
	
	\bibitem{genbootstrap}
	Li T, Lawson J.
	\newblock A Generalized Bootstrap Procedure of the Standard Error and
	Confidence Interval Estimation for Inverse Probability of Treatment
	Weighting.
	\newblock Multivariate Behavioral Research. 2023;0(0):1--15.
	\newblock \doi{10.1080/00273171.2023.2254541}.
	
	\bibitem{greenland1999}
	Greenland S, Pearl J, Robins JM.
	\newblock {Causal diagrams for epidemiologic research}.
	\newblock Epidemiology. 1999;10(1):37--48.
	
	\bibitem{spreaficoBMJOpen}
	Spreafico M, Ieva F, Arlati F, Capello F, Fatone F, Fedeli F, et~al.
	\newblock {Novel longitudinal Multiple Overall Toxicity (MOTox) score to
		quantify adverse events experienced by patients during chemotherapy
		treatment: a retrospective analysis of the MRC BO06 trial in osteosarcoma}.
	\newblock BMJ Open. 2021;11(12):e053456.
	\newblock \doi{10.1136/bmjopen-2021-053456}.
	
	\bibitem{rosen1985}
	Rosen G, Nirenberg A.
	\newblock {Neoadjuvant chemotherapy for osteogenic sarcoma: a five year
		follow-up (T-10) and preliminary report of new studies (T-12)}.
	\newblock Progress in clinical and biological research. 1985;201:39--51.
	
	\bibitem{ctcae3}
	{US Department of Health and Human Services}.: {Common Terminology Criteria for
		Adverse Events v3.0 (CTCAE)}.
	\newblock \url{https://www.eortc.be/services/doc/ctc/ctcaev3.pdf}.
	
	\bibitem{landmark2007}
	van Houwelingen HC.
	\newblock Dynamic Prediction by Landmarking in Event History Analysis.
	\newblock Scandinavian Journal of Statistics. 2007;34(1):70--85.
	\newblock \doi{https://doi.org/10.1111/j.1467-9469.2006.00529.x}.
	
	\bibitem{landmarkbook}
	van Houwelingen HC, Putter H.
	\newblock Dynamic Prediction in Clinical Survival Analysis.
	\newblock Raton B, editor. Chapman \& Hall/CRC Press; 2011.
	
	\bibitem{landmark2017}
	Putter H, van Houwelingen HC.
	\newblock {Understanding landmarking and its relation with time-dependent cox
		regression}.
	\newblock Statistics in Bioscience. 2017;9(2):489--503.
	
	\bibitem{ravani2017}
	Williamson T, Ravani P.
	\newblock {Marginal structural models in clinical research: when and how to use
		them?}
	\newblock Nephrology Dialysis Transplantation. 2017;32(suppl 2):ii84--ii90.
	\newblock \doi{10.1093/ndt/gfw341}.
	
	\bibitem{bours2021}
	Bours MJL.
	\newblock {Tutorial: A nontechnical explanation of the counterfactual
		definition of effect modification and interaction}.
	\newblock Journal of Clinical Epidemiology. 2021;134:113--124.
	\newblock \doi{10.1016/j.jclinepi.2021.01.022}.
	
	\bibitem{collins2013}
	Collins M, Wilhelm M, Conyers R, Herschtal A, Whelan J, Bielack S, et~al.
	\newblock {Benefits and Adverse Events in Younger Versus Older Patients
		Receiving Neoadjuvant Chemotherapy for Osteosarcoma: Findings From a
		Meta-Analysis}.
	\newblock Journal of Clinical Oncology. 2013;31(18):2303--2312.
	\newblock \doi{10.1200/JCO.2012.43.8598}.
	
	\bibitem{weinberg2007}
	Weinberg CR.
	\newblock {Can {DAGs} clarify effect modification?}
	\newblock Epidemiology. 2007;18(5):569--572.
	\newblock \doi{10.1097/EDE.0b013e318126c11d}.
	
	\bibitem{attia2022}
	Attia J, Holliday E, Oldmeadow C.
	\newblock {A proposal for capturing interaction and effect modification using
		DAGs}.
	\newblock International Journal of Epidemiology. 2022 06;51(4):1047--1053.
	\newblock \doi{10.1093/ije/dyac126}.
	
	\bibitem{hernan2000epi}
	Hernán MA, Brumback B, Robins JM.
	\newblock {Marginal Structural Models to Estimate the Causal Effect of
		Zidovudine on the Survival of HIV-Positive Men}.
	\newblock Epidemiology. 2000;11(5):561--570.
	\newblock \doi{10.1097/00001648-200009000-00012}.
	
	\bibitem{robins2000}
	Robins JM, Hernán MA, Brumback B.
	\newblock {Marginal structural models and causal inference in epidemiology}.
	\newblock Epidemiology. 2000;11(5):550--560.
	\newblock \doi{10.1097/00001648-200009000-00011}.
	
	\bibitem{Binder1992}
	Binder DA.
	\newblock Fitting Cox's proportional hazards models to survey data.
	\newblock Biometrika. 1992;79:139--147.
	\newblock \doi{10.2307/2337154}.
	
	\bibitem{Lin2000}
	Lin DY.
	\newblock {On fitting Cox's proportional hazards models to survey data}.
	\newblock Biometrika. 2000;87:37--47.
	
	\bibitem{cole2009epi}
	Cole SR, Frangakis CE.
	\newblock {The Consistency Statement in Causal Inference}.
	\newblock Epidemiology. 2009;20(3-5).
	\newblock \doi{10.1097/EDE.0b013e31818ef366}.
	
	\bibitem{austinstuart2015}
	Austin PC, Stuart EA.
	\newblock {Moving towards best practice when using inverse probability of
		treatment weighting (IPTW) using the propensity score to estimate causal
		treatment effects in observational studies}.
	\newblock Statistics in Medicine. 2015;34(28):3661--3679.
	\newblock \doi{10.1002/sim.6607}.
	
	\bibitem{Royston2011}
	Royston P, Parmar MKB.
	\newblock The use of restricted mean survival time to estimate the treatment
	effect in randomized clinical trials when the proportional hazards assumption
	is in doubt.
	\newblock Statistics in Medicine. 2011;30(19):2409--2421.
	\newblock \doi{10.1002/sim.4274}.
	
	\bibitem{Uno2014}
	Uno H, Claggett B, Tian L, Inoue E, Gallo P, Miyata T, et~al.
	\newblock Moving beyond the hazard ratio in quantifying the between-group
	difference in survival analysis.
	\newblock Journal of Clinical Oncology. 2014;32(22):2380--2385.
	\newblock \doi{10.1200/JCO.2014.55.2208}.
	
	\bibitem{R}
	{R Core Team}.: {R: A Language and Environment for Statistical Computing}.
	\newblock Vienna, Austria.
	\newblock \url{https://www.R-project.org/}.
	
	\bibitem{ipw}
	{van der Wal} WM, Geskus RB.
	\newblock {ipw}: An {R} Package for Inverse Probability Weighting.
	\newblock Journal of Statistical Software. 2011;43(13):1--23.
	\newblock \doi{10.18637/jss.v043.i13}.
	
	\bibitem{survival}
	Therneau T.: {A Package for Survival Analysis in R}.
	\newblock \url{https://www.R-project.org/package=survival}.
	
	\bibitem{schemper1996}
	Schemper M, Smith TL.
	\newblock {A note on quantifying follow-up in studies of failure time}.
	\newblock Controlled Clinical Trials. 1996;17(4):343--346.
	\newblock \doi{10.1016/0197-2456(96)00075-x}.
		
	\bibitem{colehernan2008}
	Cole SR, Hern\'{a}n MA.
	\newblock {Constructing Inverse Probability Weights for Marginal Structural
		Models}.
	\newblock American Journal of Epidemiology. 2008;168(6):656–664.
	\newblock \doi{10.1093/aje/kwn164}.
	
	\bibitem{karim2014}
	Karim ME, Gustafson P, Petkau J, Yinshan~Zhao AS, Kingwell E, Evans C, et~al.
	\newblock {Marginal Structural Cox Models for Estimating the Association
		Between $\beta$-Interferon Exposure and Disease Progression in a Multiple
		Sclerosis Cohort}.
	\newblock American Journal of Epidemiology. 2014;180(2):160--171.
	\newblock \doi{10.1093/aje/kwu125}.
	
	\bibitem{karim2017}
	Karim ME, Petkau J, Gustafson P, Tremlett H, Group TBS.
	\newblock {On the application of statistical learning approaches to construct
		inverse probability weights in marginal structural Cox models: Hedging
		against weight-model misspecification}.
	\newblock Communications in Statistics - Simulation and Computation.
	2017;46(10):7668--7697.
	\newblock \doi{10.1080/03610918.2016.1248574}.

	
	
\end{thebibliography}
\end{document}